\documentclass{article}

\usepackage{amsthm}
\usepackage{amsmath}
\usepackage{natbib}
\usepackage[colorlinks,citecolor=blue,urlcolor=blue,filecolor=blue,backref=page]{hyperref}
\usepackage{graphicx}

\usepackage{algorithm}
\usepackage{algorithmic}
\usepackage{theorem} 
\theoremstyle{break}
\usepackage{bm}
\usepackage{bbm}
\usepackage{lscape}
\usepackage{multirow} 

\begin{document}

\title{Bayes factors for partially observed stochastic epidemic models}
\author{Muteb Alharthi\footnote{Taif University, Saudi Arabia, {\tt muteb.faraj@tu.edu.sa}} \and Theodore Kypraios\footnote{University of Nottingham, United Kingdom {\tt theodore.kypraios@nottingham.ac.uk}} \and Philip D. O'Neill\footnote{University of Nottingham, United Kingdom {\tt philip.oneill@nottingham.ac.uk}}}
\maketitle

\begin{abstract}
We consider the problem of model choice for stochastic epidemic models given partial observation of a disease outbreak through time.
Our main focus is on the use of Bayes factors. Although Bayes factors have appeared in the epidemic modelling literature before, they
can be hard to compute and little attention
has been given to fundamental questions concerning their utility. In this paper we derive analytic expressions for Bayes factors given complete
observation through time, which suggest practical guidelines for model choice problems. We extend the power posterior method for computing Bayes
factors so as to account for missing data and apply this approach to partially observed epidemics. For comparison, we also explore the use of
a deviance information criterion for missing data scenarios. The methods are illustrated via examples involving both simulated and real data.
\end{abstract}

\section{Introduction}
This paper is concerned with the problem of choosing between a small number of competing infectious disease transmission models, given partial observation
of an epidemic outbreak through time. A key reason to consider such a problem is that the models represent different hypotheses about disease
spread, such as the infection mechanism, the nature of the contact structure between individuals in the population, or other disease characteristics.
A secondary reason to compare models is that they are often used to simulate potential future outbreaks, perhaps with a view to designing control strategies,
in which case it is computationally more efficient to employ the simplest possible model which can represent observed data reasonably well. Note that it is typically
the case that we only have a few models under consideration, in contrast to variable-selection problems that occur in regression modelling. Also, our
focus here is not on the assessment of model fit, itself concerned with whether or not a single specific epidemic model adequately describes the data to hand.

Within the epidemic modelling literature, there is to date no definitively preferred method for model choice. In the Bayesian setting, approaches include the use
of Bayes factors, \citep[e.g.][]{neal2004, oneill2005}, criteria such as the Deviance Information Criterion (DIC) \citep[e.g.][]{worby2013, lau2014, deeth2015}, and methods based on the predictive distribution of future outbreaks \citep{zhang2014}. Here we focus on the use of Bayes factors. The most common approach to calculating Bayes factors for epidemics has been to use reversible jump Markov chain Monte Carlo methods \citep{neal2004, oneill2005}, although these are often problematic in
practice due to the challenge of designing efficient algorithms.
\cite{panayiota2015} used a combination of MCMC methods and importance sampling to estimate marginal likelihoods, from which Bayes factors can be calculated.
\cite{knock2014} used a path-sampling method \citep{gelman1998} to compare epidemic models given data on the final outcome of the epidemic. This is somewhat related to the methods we develop, albeit for different kinds of data.

In this paper we extend the power posterior method for calculating Bayes factors \citep{friel2008, friel2014} to a missing-data situation that commonly occurs in
epidemic modelling. In addition to this computational method, we also derive analytic expressions for Bayes factors in the setting where an epidemic outbreak is
completely observed. Although such detailed observation is not that common in practice, our results are of theoretical interest and also provide practical insight into the choice and influence of prior distributions for parameters of the competing models. For comparison, we also consider a form of DIC suitable for missing data.

Throughout the paper we focus on the Susceptible-Infective-Removed (SIR) epidemic model, the most widely-studied stochastic epidemic model. However, the computational
methods we develop could be applied to more complex models. For illustration we consider two specific kinds of model choice question: one in which models with different
infectious period distributions are compared, and one in which models with different infection mechanisms are compared. Again, other comparisons are possible using
our methods.

The paper is arranged as follows. Section 2 recalls preliminary information on epidemic models, Bayes factors, DIC methods and power posterior methods, extending
the latter to a missing-data situation. In Section 3 we derive analytical expressions for Bayes factors given completely observed outbreaks, and in Section 4 we describe computational methods for partially observed outbreaks. Section 5 contains illustrative examples of the methods, and concluding comments are found in Section 6.

\section{Preliminaries}\label{Preliminaries}
\subsection{The stochastic SIR epidemic model}
The stochastic SIR epidemic model is defined as follows \citep[see e.g.][]{andersson2000}. Consider a closed population of $N$ individuals. At any time, each
individual is either susceptible, infective or removed. Susceptible individuals have not contracted the disease but are able to do so. Infective individuals have
the disease and can pass it on to others. Removed individuals are no longer infective and play no further part in disease spread. In applications, the removed state
is context-specific and could correspond to immunity, isolation, death, or similar outcomes. The population initially consists of $n$ susceptible individuals and $a$
infectives who have just become infective. Each infective individual remains so for a period of time, called the infectious period, which is drawn from a specified non-negative probability distribution $T_I$. At the end of its infectious period an individual is immediately removed. During its infectious period, a given infective
has contacts with a given susceptible in the population at times given by the points of a Poisson process of rate $\beta n^{-1}$. The first such contact, if it occurs,
results in the susceptible immediately becoming infective, and later contacts have no effect. All infectious periods and all Poisson processes are mutually independent.
The epidemic ends as soon as there are no more infectives remaining in the population.

For any time $t$, denote by $X(t)$ and $Y(t)$ respectively the numbers of susceptible and infective individuals currently in the population. It follows that infections occur in
the population according to a Poisson process of rate $\beta n^{-1} X(t) Y(t)$. We will also consider a variant of the SIR model in which the overall infection rate is $\beta n^{-1} X(t) Y^p(t)$ for $p \in (0,1)$. Such models were first introduced by \cite{severo1969} and relax the assumption that the overall
infection rate increases linearly with $Y(t)$ \citep[see also][and references therein]{oneill.non2012}. We will focus on two particular choices of infectious period distribution, namely exponential and gamma, both of which frequently appear in the epidemic modelling literature. Finally, we will assume throughout that there is one initial infective, although this constraint can easily be relaxed.

\subsection{Bayes factors}
Suppose we have two competing epidemic models, $m_1$ and $m_2$, with parameters $\bm{\theta}_1$ and $\bm{\theta}_2$, respectively, and that we have observed data $\bm{y}$.
The Bayes factor for $m_1$ relative to $m_2$ is defined by
\[
BF_{12} = \frac{\pi( \bm{y} | m_1 )}{\pi(\bm{y} | m_2)} = \frac{ \int \pi(\bm{y} |\bm{\theta}_1) \pi(\bm{\theta}_1) d \bm{\theta}_1}{ \int \pi( \bm{y} |\bm{\theta}_2) \pi(\bm{\theta}_2) d \bm{\theta}_2}
\]
where here, and throughout the paper, $\pi$ denotes a probability mass or density function, as appropriate. For partially observed epidemic models, the likelihood
term $\pi( \bm{y} | \bm{\theta})$ is typically intractable, and a common approach is to introduce auxiliary variables, $\bm{x}$ say, such that the augmented likelihood
$\pi(\bm{y}, \bm{x} | \bm{\theta})$ is available in closed form and can be computed efficiently. Estimation of the posterior distribution of $\bm{\theta}$ can then be achieved via Markov chain Monte Carlo methods \citep{gibson1998, oneill1999}. In most situations, $\bm{x}$ will describe the infection process, since this is usually unobserved.

The computation of Bayes factors is, in general, a challenging problem. Many approaches exist, but here we focus specifically on the power posterior method. One
attractive aspect of this approach is that it is relatively prescriptive, meaning that the user does not have that many implementation choices which could seriously
affect the performance of the resulting algorithm. Also, as we now explain, the method can be extended to cover the kind of missing data problem that usually arises
in the context of epidemic modelling.

\subsection{The power posterior method for models incorporating missing data} \label{Power posteriors for models incorporating missing data}
We now extend the power posterior approach \citep{lartillot2006,friel2008,friel2014} to models incorporating missing data. The method itself provides a way of
calculating the marginal density $\pi( \bm{y} |m_i)$, $i=1, 2$, from which the Bayes factor can be obtained.

Let $\bm{y}$ and $\bm{x}$ denote the observed and the missing data respectively with $\bm{\theta}$ representing the model parameters. We refer to $(\bm{y}, \bm{x})$ as the complete data. Note that in the epidemic settings that we will consider, neither $\pi(\bm{y}\vert \bm{\theta})$ nor $\pi(\bm{y}\vert \bm{x}, \bm{\theta})$ are typically tractable, but we can compute the augmented likelihood function $\pi(\bm{y}, \bm{x} \vert \bm{\theta})$.

For $t\in[0, 1]$, we define the power posterior for the missing data scenario as
\[
\pi_{t}(\bm{\theta}, \bm{x}\vert\bm{y}) \propto \pi(\bm{y}, \bm{x}\vert \bm{\theta})^{t} \pi(\bm{\theta}),
\]
with the normalizing constant
\[
z_{t}(\bm{y}) =\int_{\bm{x}} \int_{\bm{\theta}} \pi(\bm{y}, \bm{x}\vert \bm{\theta})^{t} \pi(\bm{\theta}) d\bm{\theta}\; d\bm{x}.
\]
Thus, noting that $z_{t=1}(\bm{y}) = \pi(\bm{y})$ and $z_{t=0}(\bm{y}) = 1$,
\begin{equation}
\log(\pi(\bm{y}))= \log \left[\frac{z_{t=1}(\bm{y})}{z_{t=0}(\bm{y})} \right]= \int_{0}^{1} E_{\bm{\theta}, \bm{x}\vert\bm{y}, t} \log \left[\pi(\bm{y}, \bm{x}\vert \bm{\theta})\right] dt ,
\label{z_eqn}
\end{equation}
where the second equality in (\ref{z_eqn}) can be derived by adapting the arguments of \cite{lartillot2006}, as follows.
\begin{eqnarray*}
\frac{d}{dt}\log(z_{t}(\bm{y}))&=& \frac{1}{z_{t}(\bm{y})}\frac{d}{dt} z_{t}(\bm{y})\nonumber\\
&=& \frac{1}{z_{t}(\bm{y})} \int_{\bm{x}} \int_{\bm{\theta}} \frac{d}{dt} \pi(\bm{y}, \bm{x}\vert \bm{\theta})^{t} \pi(\bm{\theta}) d\bm{\theta}\; d\bm{x} \nonumber\\
&=& \frac{1}{z_{t}(\bm{y})} \int_{\bm{x}} \int_{\bm{\theta}} \pi(\bm{y}, \bm{x}\vert \bm{\theta})^{t} \log \left[\pi(\bm{y}, \bm{x}\vert \bm{\theta})\right] \pi(\bm{\theta}) d\bm{\theta}\; d\bm{x} \nonumber\\
&=& \int_{\bm{x}} \int_{\bm{\theta}} \frac{\pi(\bm{y}, \bm{x}\vert \bm{\theta})^{t} \pi(\bm{\theta})}{z_{t}(\bm{y})}  \log \left[\pi(\bm{y}, \bm{x}\vert \bm{\theta})\right]  d\bm{\theta}\; d\bm{x} \nonumber\\
&=& \int_{\bm{x}} \int_{\bm{\theta}}  \log \left[\pi(\bm{y}, \bm{x}\vert \bm{\theta})\right] \pi_{t}(\bm{\theta}, \bm{x}\vert\bm{y})  d\bm{\theta}\; d\bm{x} \nonumber\\
&=& E_{\bm{\theta}, \bm{x}\vert\bm{y}, t} \log \left[\pi(\bm{y}, \bm{x}\vert \bm{\theta})\right] .\nonumber\\
\end{eqnarray*}
Note that the above argument requires a regularity condition, specifically permitting the exchange of order of integration and differentiation.

By integrating with respect to $t\in[0, 1]$, we have
\begin{eqnarray*}
\log(\pi(\bm{y}))&=& \log(z_{t=1}(\bm{y}))- \log(z_{t=0}(\bm{y}))\\
&=& \log(z_{t=1}(\bm{y}))\\
&=& \int_{0}^{1} E_{\bm{\theta}, \bm{x}\vert\bm{y}, t} \log \left[\pi(\bm{y}, \bm{x}\vert \bm{\theta})\right] dt .
\end{eqnarray*}

We shall evaluate the final integral numerically, by evaluating it at a finite number of $t$ values, namely $0 = t_0 < t_1 < \ldots < t_r=1$. To reduce the resulting approximation error, we follow \cite{friel2014} and make use of the fact that the gradient of the expected log-likelihood curve equals its variance. Specifically,
\begin{eqnarray*}
\frac{d}{dt}E_{\bm{\theta}, \bm{x}\vert\bm{y}, t} \log \left[\pi(\bm{y}, \bm{x}\vert \bm{\theta})\right]&=& \frac{d}{dt} \int_{\bm{x}} \int_{\bm{\theta}}  \log \left[\pi(\bm{y}, \bm{x}\vert \bm{\theta})\right] \pi_{t}(\bm{\theta}, \bm{x}\vert\bm{y})  d\bm{\theta}\; d\bm{x}\\
&=&\int_{\bm{x}} \int_{\bm{\theta}} \log \left[\pi(\bm{y}, \bm{x}\vert \bm{\theta})\right] \frac{d}{dt} \pi_{t}(\bm{\theta}, \bm{x}\vert\bm{y})  d\bm{\theta}\; d\bm{x},
\end{eqnarray*}
where

\begin{eqnarray*}
 \frac{d}{dt} \pi_{t}(\bm{\theta}, \bm{x}\vert\bm{y}) &=& \frac{d}{dt} \left[\frac{\pi(\bm{y}, \bm{x}\vert \bm{\theta})^{t} \pi(\bm{\theta})}{z_{t}(\bm{y})} \right]\\
 &=& \frac{z_{t}(\bm{y}) \pi(\bm{y}, \bm{x}\vert \bm{\theta})^{t} \pi(\bm{\theta}) \log \left[\pi(\bm{y}, \bm{x}\vert \bm{\theta})\right] - \pi(\bm{y}, \bm{x}\vert \bm{\theta})^{t} \pi(\bm{\theta}) \frac{d}{dt} z_{t}(\bm{y}) }{z^2_{t}(\bm{y})}\\
 &=& \frac{\pi(\bm{y}, \bm{x}\vert \bm{\theta})^{t} \pi(\bm{\theta})}{z_{t}(\bm{y})} \left[ \log \left[\pi(\bm{y}, \bm{x}\vert \bm{\theta})\right] - \frac{1}{z_{t}(\bm{y})}\frac{d}{dt} z_{t}(\bm{y}) \right]\\
 &=& \pi_{t}(\bm{\theta}, \bm{x}\vert\bm{y}) \left[\log \left[\pi(\bm{y}, \bm{x}\vert \bm{\theta})\right]- \frac{d}{dt}\log(z_{t}(\bm{y}))\right].
\end{eqnarray*}
Hence,
\begin{eqnarray*}
\frac{d}{dt}E_{\bm{\theta}, \bm{x}\vert\bm{y}, t} \log \left[\pi(\bm{y}, \bm{x}\vert \bm{\theta})\right]&=& \int_{\bm{x}} \int_{\bm{\theta}}  \left(\log \left[\pi(\bm{y}, \bm{x}\vert \bm{\theta})\right]\right)^2 \pi_{t}(\bm{\theta}, \bm{x}\vert\bm{y})  d\bm{\theta}\; d\bm{x}\\
&-& \frac{d}{dt}\log(z_{t}(\bm{y})) \int_{\bm{x}} \int_{\bm{\theta}}  \log \left[\pi(\bm{y}, \bm{x}\vert \bm{\theta})\right] \pi_{t}(\bm{\theta}, \bm{x}\vert\bm{y})  d\bm{\theta}\; d\bm{x}\\
&=& E_{\bm{\theta}, \bm{x}\vert\bm{y}, t} \left(\log \left[\pi(\bm{y}, \bm{x}\vert \bm{\theta})\right]\right)^2 - \left( E_{\bm{\theta}, \bm{x}\vert\bm{y}, t} \log \left[\pi(\bm{y}, \bm{x}\vert \bm{\theta})\right]\right)^2\\
&=& V_{\bm{\theta}, \bm{x}\vert\bm{y}, t}  \log \left[\pi(\bm{y}, \bm{x}\vert \bm{\theta})\right].
\end{eqnarray*}
Using the corrected trapezoidal rule form \citep{atkinson2004}, namely
\begin{equation*}
\int_{a}^{b} f(y) dy \approx \frac{(b-a)}{2}\left[f(a)+f(b)\right] - \frac{(b-a)^2}{12} \left[f^{\prime}(b) - f^{\prime}(a)\right],
\end{equation*}
we have the following extended power posterior estimate of $\log(\pi(\bm{y}))$ :
\begin{eqnarray} \label{improvepp missing eq}
\log(\pi(\bm{y}))&\approx& \sum_{j=1}^{r} \frac{1}{2}(t_j - t_{j-1})\nonumber\\
&\times& \left\{E_{\bm{\theta}, \bm{x}\vert\bm{y}, t_j} \log \left[\pi(\bm{y}, \bm{x}\vert \bm{\theta})\right] + E_{\bm{\theta}, \bm{x}\vert\bm{y}, t_{j-1}} \log \left[\pi(\bm{y}, \bm{x}\vert \bm{\theta})\right] \right\}\nonumber\\
&-& \sum_{j=1}^{r} \frac{1}{12}(t_j - t_{j-1})^2 \nonumber\\ &\times& \left\{V_{\bm{\theta}, \bm{x}\vert\bm{y}, t_j} \log \left[\pi(\bm{y}, \bm{x}\vert \bm{\theta})\right] - V_{\bm{\theta}, \bm{x}\vert\bm{y}, t_{j-1}} \log \left[\pi(\bm{y}, \bm{x}\vert \bm{\theta})\right] \right\}. \nonumber\\
& &
\end{eqnarray}

Algorithm \ref{algor1} can be used to implement the extended power posterior method for missing data models.
We follow the recommendation of \cite{friel2008} for the choice of $t_j$ values in (\ref{improvepp missing eq}). This choice ensures that many of the $t_j$ values are close to zero, where the expected log-likelihood curve often changes rapidly in practice. The $t_j$ values are often called temperatures, and the collection
of values called a temperature ladder, this terminology arising because the power posterior method is a form of so-called thermodynamic integration.

\begin{algorithm} [h!t]
\caption{MCMC algorithm for estimating the marginal likelihood via the missing-data power posterior approach} \label{algor1}
\begin{algorithmic}
\STATE 1. Initialise algorithm with $\bm{x}^0$ and $\bm{\theta}^0$.
\vskip 3pt
\STATE 2. For $j = 0,..., r$:
\vskip 2pt
\STATE a. Set $t_j=(j/r)^c$, where $c>1$ is a constant.
\vskip 1pt
\STATE b. Generate a sample $\{(\bm{\theta}^{(1)}_j, \bm{x}^{(1)}_j), ..., (\bm{\theta}^{(M)}_j, \bm{x}^{(M)}_j) \}$ from $\pi_{t_j}(\bm{\theta}, \bm{x} \vert\bm{y})$ via an MCMC sampling scheme.
\vskip 1pt
\STATE c. Estimate $E_{\bm{\theta}, \bm{x}\vert\bm{y}, t_j} \log \left[\pi(\bm{y}, \bm{x}\vert \bm{\theta})\right]$ and $V_{\bm{\theta}, \bm{x}\vert\bm{y}, t_j} \log \left[\pi(\bm{y}, \bm{x}\vert \bm{\theta})\right]$ using the sample from b.
\vskip 1pt
\STATE d. While $j < r$, initialise the next chain at the previous posterior mean of $\pi_{t_j}(\bm{\theta}\vert\bm{y}, \bm{x})$.
\vskip 3pt
\STATE 3. Estimate $\log(\pi(\bm{y}))$ using (\ref{improvepp missing eq}).
\end{algorithmic}
\end{algorithm}

\subsection{DIC for models with missing data} \label{DIC for models with missing data}
Although Bayes factors are our primary focus, for comparison we will also compute a form of DIC suitable for missing data situations.
\cite{celeux2006} propose various options; the one best-suited to our setting, in the sense that it is suitable for situations where the missing data
are not our main focus, and we can compute it, is
\begin{equation}
 \mbox{DIC}_6 = -4E_{\bm{\theta}, \bm{x}\vert\bm{y}}[\log(\pi(\bm{y}, \bm{x}\vert\bm{\theta}))] + 2 E_{\bm{x}\vert\bm{y},\hat{\bm{\theta}}}[\log(\pi(\bm{y},  \bm{x}\vert\hat{\bm{\theta}}))].
 \label{DIC_formula}
\end{equation}
Calculation of this quantity requires two runs of an MCMC algorithm. The first is to derive $E_{\bm{\theta}, \bm{x}\vert\bm{y}}[\log(\pi(\bm{y}, \bm{x}\vert\bm{\theta}))]$ and $E_{\bm{\theta}\vert \bm{y}} (\bm{\theta}\vert \bm{y})  = \hat{\bm{\theta}}$, and the second run is to obtain $E_{\bm{x}\vert\bm{y},\hat{\bm{\theta}}}[\log(\pi(\bm{y},  \bm{x}\vert\hat{\bm{\theta}}))]$ setting $\bm{\theta}= \hat{\bm{\theta}}$ and allowing $\bm{x}$ to vary.
In our epidemic setting, $\hat{\bm{\theta}}$ contains posterior point estimates of the model parameters, the identity of the initial infective and the initial
infection time. The preferred model from those under consideration is the one with the lowest DIC$_6$ value.

\section{Model selection given complete outbreak data}\label{Model selection given complete outbreak data}
In this section we show that Bayes factors for the epidemic models of interest can be computed explicitly if complete data are available. This situation is rare in
practice, although it can arise when outbreaks are being closely monitored (e.g. in the early stages of a suspected major epidemic, or in experimental settings for
animal diseases). Nevertheless, we gain some insight into the value of Bayes factors as a tool for model choice, particularly with respect to the choice of within-model prior distribution.

\subsection{The SIR model with different infectious periods}\label{The SIR model with different infectious periods}
Suppose we observe an epidemic among a population of $N$ individuals of whom initially $n$ are susceptible and one is infective. Denote by $n_R$ the total
number of individuals ever infected, including the initial infective, and label these $n_R$ individuals $1, \ldots, n_R$. The remaining individuals are
labelled $n_R+1, \ldots, N$. For $j =1, \ldots, N$ let $I_j$ and $R_j$ denote, respectively, the infection and removal time of individual $j$, with
$I_j = R_j = \infty$ for $j > n_R$. Let $z$ denote the label of the initial infective, so that $I_z < I_j$, for all $j\neq z$.
Finally let $\bm{I}= (I_1, \ldots, I_{z-1}, I_{z+1}, \ldots, I_{n_R})$ denote the vector of infection times of infected individuals other than the initial infective,
and let $\bm{R}= (R_1,...,R_{n_R})$ denote the vector of removal times of all infected individuals.

We consider two competing SIR models with identical infection mechanisms but different choices of infectious period distribution $T_I$. Specifically,
model $m_1$ has $T_I\sim \mbox{Exp}(\gamma)$ and model $m_2$ has $T_I\sim \mbox{Gamma}(\alpha, \delta)$ with shape parameter $\alpha$ assumed known.
The likelihoods of $(\bm{I}, \bm{R})$ under the two models are
\begin{equation*}
\pi(\bm{I},\bm{R}\vert\beta, \gamma, I_z, z, m_1)= \beta^{n_R-1}\prod_{j=1,j\neq z}^{n_R} n^{-1}Y(I_j-) \times e^{ -\beta n^{-1} A} \times\gamma^{n_R}\; e^{-\gamma\sum_{j=1}^{n_R}(R_j-I_j)}
\end{equation*}
and
\begin{eqnarray*}
\pi(\bm{I},\bm{R}\vert \alpha, \beta, \delta, I_z, z, m_2)&=& \beta^{n_R-1}\prod_{j=1,j\neq z}^{n_R} n^{-1}Y(I_j-) \times e^{ -\beta n^{-1} A} \\
&\times& \Gamma^{-n_R}(\alpha)\times \prod_{j=1}^{n_R}(R_j-I_j)^{\alpha-1}\times \delta^{\alpha n_R}\; e^{-\delta \sum_{j=1}^{n_R}(R_j-I_j)},
\end{eqnarray*}
where
\begin{equation*}
A= \int_{I_z}^{R_{n_R}}X(t)Y(t) dt = \sum_{j=1}^{n_R}\sum_{k=1}^N( R_j \wedge I_k - I_k \wedge I_j),
\end{equation*}
$\Gamma^{-n_R}(\alpha)$ denotes $(\Gamma(\alpha))^{-n_R}$ and $Y(t-) = \lim_{s \uparrow t} Y(s)$, see e.g. \cite{kypraios2007}.

By assigning an independent gamma prior distribution for each of the model parameters, namely $\mbox{Gamma}(\lambda_\zeta, \nu_\zeta)$, where $\zeta=\beta,\;\gamma,\;\delta$, the Bayes factor can be derived explicitly in this case as follows.
\begin{eqnarray} \label{true.BFexpgam}
BF_{12} &=& \frac{\pi(\bm{I}, \bm{R}\vert m_1)}{\pi(\bm{I}, \bm{R}\vert m_2)}
= \frac{\int_{\gamma}\int_{\beta}\pi(\bm{I}, \bm{R}\vert \beta, \gamma) \pi(\beta)\pi(\gamma) d\beta\;d\gamma}{\int_{\delta}\int_{\beta}\pi(\bm{I}, \bm{R}\vert \beta, \alpha, \delta) \pi(\beta) \pi(\delta) d\beta \;d\delta} \nonumber\\
&=& \frac{\nu^{\lambda_\beta}_\beta\; \nu^{\lambda_\gamma}_\gamma \;\Gamma(\lambda_\beta)\;\Gamma(\lambda_\delta)}{\nu^{\lambda_\beta}_\beta\; \nu^{\lambda_\delta}_\delta \; \Gamma(\lambda_\beta)\;\Gamma(\lambda_\gamma)}\times\frac{\Gamma^{n_R}(\alpha)\times\prod_{j=1,j\neq z}^{n_R} n^{-1}Y(I_j-)}{\prod_{j=1,j\neq z}^{n_R} n^{-1}Y(I_j-) \times \prod_{j=1}^{n_R}(R_j-I_j)^{\alpha-1}}\nonumber\\
&\times& \frac{\int_{\beta} \beta^{n_R-1} \times e^{ -\beta n^{-1} A} \times \beta^{\lambda_\beta -1} e^{-\nu_\beta \beta} d\beta}
{\int_{\beta} \beta^{n_R-1} \times e^{ -\beta n^{-1} A} \times \beta^{\lambda_\beta -1} e^{-\nu_\beta \beta} d\beta}\nonumber\\
&\times& \frac{\int_{\gamma} \gamma^{n_R}\; e^{-\gamma\sum_{j=1}^{n_R}(R_j-I_j)}\times \gamma^{\lambda_\gamma -1} e^{-\nu_\gamma \gamma} d\gamma}
{\int_{\delta} \delta^{\alpha n_R}\; e^{-\delta \sum_{j=1}^{n_R}(R_j-I_j)} \times \delta^{\lambda_\delta -1} e^{-\nu_\delta \delta} d\delta}\nonumber\\
&=& \frac{\nu^{\lambda_\gamma}_\gamma \;\Gamma(\lambda_\delta)\times\Gamma^{n_R}(\alpha)}{\nu^{\lambda_\delta}_\delta \; \Gamma(\lambda_\gamma)\times\prod_{j=1}^{n_R}(R_j-I_j)^{\alpha-1}} \nonumber\\ &\times& \frac{(\nu_\delta+\sum_{j=1}^{n_R}(R_j-I_j))^{ \alpha n_R+\lambda_\delta}\times\Gamma(n_R+\lambda_\gamma)}
{(\nu_\gamma+\sum_{j=1}^{n_R}(R_j-I_j))^{n_R+\lambda_\gamma}\times\Gamma(\alpha n_R +\lambda_\delta)}.
\end{eqnarray}

The resulting Bayes factor is independent of the infection rate prior parameters. This is a consequence of the fact that the likelihood expressions
can be factorized into parts corresponding to the infection and removal processes, and the former are the same in both models. Note also that the Bayes
factor is identical to that obtained by comparing exponential and gamma distributions for a sequence of independent and identically distributed observations
$R_1 - I_1, \ldots, R_{n_R} - I_{n_R}$, although in our setting things are slightly different because the observations and $n_R$ are not independent.

In the epidemic modelling literature, it is often the case that prior parameters for positive quantities such as rate parameters are assigned the same vague prior
distributions. Here, this assumption gives $\lambda_\gamma=\lambda_\delta=\lambda$ and $\nu_\gamma=\nu_\delta=\nu$, where $\lambda \geq 1$ and $\nu$ is a small positive number. The Bayes factor in (\ref{true.BFexpgam}) then becomes
\begin{equation} \label{BFlimitexpgam1}
BF_{12} = \frac{\Gamma(n_R+\lambda)\;\;\Gamma^{n_R}(\alpha)}{ \Gamma( \alpha n_R+\lambda)\;\prod_{j=1}^{n_R}(R_j-I_j)^{\alpha-1}} \times \left(\nu + \sum_{j=1}^{n_R}(R_j-I_j)\right)^{n_R(\alpha-1)} .
\end{equation}
Reformulating (\ref{BFlimitexpgam1}) in terms of the mean and variance of the prior distribution yields
\[
BF_{12} = \frac{\Gamma\left(n_R+\frac{\mu^2}{\sigma^2}\right)\;\;\Gamma^{n_R}(\alpha)}{ \Gamma\left(\alpha n_R+\frac{\mu^2}{\sigma^2}\right)\;\prod_{j=1}^{n_R}(R_j-I_j)^{\alpha-1}} \times \left(\frac{\mu}{\sigma^2} + \sum_{j=1}^{n_R}(R_j-I_j)\right)^{n_R(\alpha-1)},
\]
where $\mu = \lambda/\nu$ and $\sigma^2 = \lambda/\nu^2$. Thus as $\sigma^2\rightarrow\infty$, the prior becomes increasingly diffuse and $BF_{12}$ converges to
its lower limit, that is
\begin{equation}\label{limitBF121}
BF_{12} \rightarrow \frac{\Gamma(n_R)\;\;\Gamma^{n_R}(\alpha)}{ \Gamma(\alpha n_R)\;\prod_{j=1}^{n_R}(R_j-I_j)^{\alpha-1}} \times \left(\sum_{j=1}^{n_R}(R_j-I_j)\right)^{n_R(\alpha-1)} .
\end{equation}
However, as $\sigma^2 \rightarrow 0$ the prior gets increasingly concentrated at $\mu$, and the Bayes factor becomes more decisive in supporting $m_1$, that is $BF_{12} \rightarrow \infty$.

These results suggest that diffuse priors are more appropriate. Table \ref{BF_inf_periods} shows the results of a simulation exercise in which
data sets were simulated under either $m_1$ or $m_2$, and in each case $BF_{12}$ was calculated using (\ref{limitBF121}). The resulting mean values, and the proportion
of times that $m_1$ was favoured, are presented. It can be seen that the Bayes factor discriminates effectively between the two models, even in the relatively small
population sizes of $N=30$ and $N=50$. As expected, increasing $\beta$ increases the outbreak size which in turn makes the comparison more decisive.

\begin{table}
\label{BF_inf_periods}
\begin{tabular}{lllllll}
True & & & \multicolumn{2}{c}{$E[\log BF_{12} ]$} &  \multicolumn{2}{c}{$P(BF_{12} > 1)$} \\
model & $\alpha$ &  $\beta$    & $N=30$ & $N=50$ & $N=30$ & $N=50$\\
\hline
$m_1$ & 10 & 1.5 & 42.6 & 69.5 & 0.92 & 0.92\\
$m_1$ & 5 & 1.5 & 15.8 & 23.3 & 0.84 & 0.87\\
$m_1$ & 2 & 1.5 & 1.8  & 3.2 & 0.65 & 0.67 \\
$m_1$ & 10 & 2.0 & 62.6 & 107.2 & 0.92 & 0.94\\
$m_1$ & 5 & 2.0 & 22.5 & 39.3 & 0.91 & 0.91\\
$m_1$ & 2 & 2.0 & 2.9 & 4.9 & 0.75 & 0.83\\
$m_2$ & 10 & 1.5 & -9.4 & -14.6 & 0.02 & 0.02\\
$m_2$ & 5 & 1.5 & -5.7 & -8.9 & 0.04 & 0.04\\
$m_2$ & 2 & 1.5 & -1.5 & -2.3 & 0.15 & 0.12 \\
$m_2$ & 10 & 2.0 & -15.0 & -25.3 & 0.01 & 0.01\\
$m_2$ & 5 & 2.0 & -8.4 & -14.8 & 0.02 & 0.02\\
$m_2$ & 2 & 2.0 & -2.1 & -3.7 & 0.09 & 0.07
\end{tabular}
\caption{Expected log Bayes factors for models with different infection periods, assuming diffuse prior distributions. Model $m_1$ is an SIR
model with $T_I \sim \mbox{Exp}(1)$ while $m_2$ has $T_I\sim \mbox{Gamma}(\alpha, \alpha)$, so that both models have the same mean infectious period.
Each row gives parameter values and results from 1000 simulated epidemics from the true model in which at least one new infection occurred.}
\end{table}

\subsection{The SIR model with different infection mechanisms}\label{The SIR model with different infection mechanisms}
Consider now the situation where we have two competing SIR models with different infection mechanisms but the same infectious period distribution, the latter
being essentially arbitrary. Specifically, $m_1$ is the standard SIR model in which infections occur at rate $\beta n^{-1} X(t) Y(t)$, and $m_2$ the model in which infections occur at rate $\beta n^{-1} X(t) Y^{p}(t)$. We assume $p\in(0, 0.5)$, so that $m_1$ and $m_2$ are clearly distinct, and that $p$ is known. We assume,
{\em a priori}, that (i) $\beta \sim \mbox{Gamma}(\lambda_\beta, \nu_\beta)$, and (ii) the parameters of the infectious period distribution are independent of $\beta$.

Adopting the notation and arguments of the previous section leads to
\begin{equation}
\label{true.BFstandvsmod}
BF_{12} =  \prod_{j=1, j \neq z}^{n_R} Y^{1-p}(I_j-)\times \left(\frac{\nu_\beta+n^{-1} A_{p}}{\nu_\beta+n^{-1} A}\right)^{n_R+\lambda_\beta-1},
\end{equation}
where $A=\int_{I_z}^{R_{n_R}} X(t)Y(t)\;dt$, $A_{p}=\int_{I_z}^{R_{n_R}} X(t)Y^{p}(t)\;dt$.
As expected, the Bayes factor only involves the infection process part of the likelihoods since the removal processes in the two models are assumed to be the same.
Rewriting the prior distribution for $\beta$ in terms of its mean and variance, $\sigma^2$ say, we find that

\begin{equation}\label{lowlimitBF122}
BF_{12} \rightarrow  (A_p/A)^{n_R-1}\; \prod_{j=1, j \neq z}^{n_R} Y^{1-p}(I_j-)\;\; \text{as}\;\;\sigma^2\rightarrow \infty
\end{equation}
and
\[
BF_{12} \rightarrow \prod_{j=1, j \neq z}^{n_R} Y^{1-p}(I_j-)\;\; \text{as}\;\;\sigma^2 \rightarrow 0.
\]
It is natural to suppose that the diffuse prior setting is a natural candidate for consideration. It is evident from (\ref{lowlimitBF122}) that larger $Y(I_j-)$ values in the product term will improve model discrimination; conversely, if all these values equal 1 then the product term is independent of $p$. This in turn suggests that
we require either larger or faster-growing epidemics to effectively discriminate between $m_1$ and $m_2$. This is illustrated by the results in Table \ref{BF_inf_mechanism}, which shows that we require larger values of $N$ than the infectious period distribution comparison of the previous section in order to obtain clear evidence in favour of the true model, and that increasing $\beta$ also improves discrimination. Also, as expected, as $p$ decreases then $m_1$ and $m_2$ become less similar, which also makes discrimination easier.

\begin{table}
\label{BF_inf_mechanism}
\begin{tabular}{lllllll}
True & & & \multicolumn{2}{c}{$E[\log BF_{12} ]$} &  \multicolumn{2}{c}{$P(BF_{12} > 1)$} \\
model & $p$ &  $\beta$    & $N=50$ & $N=200$ & $N=50$ & $N=200$\\
\hline
$m_1$ & 0.5 & 2.0 & 0.5 & 6.6 & 0.48 & 0.72\\
$m_1$ & 0.3 & 2.0 & 1.7 & 15.7 & 0.58 & 0.77\\
$m_1$ & 0.0 & 2.0 & 4.4 & 39.7 & 0.67 & 0.77 \\
$m_1$ & 0.5 & 4.0 & 2.9 & 18.5 & 0.77 & 0.92\\
$m_1$ & 0.3 & 4.0 & 6.1 & 41.5 & 0.85 & 0.94\\
$m_1$ & 0.0 & 4.0 & 15.0 & 98.2 & 0.92 & 0.94\\
$m_2$ & 0.5 & 2.0 & -0.8 & -1.3 & 0.09 & 0.06\\
$m_2$ & 0.3 & 2.0 & -1.2 & -1.4 & 0.07 & 0.05\\
$m_2$ & 0.0 & 2.0 & -1.2 & -1.5 & 0.06 & 0.05 \\
$m_2$ & 0.5 & 4.0 & -1.9 & -4.9 & 0.06 & 0.02\\
$m_2$ & 0.3 & 4.0 & -2.4 & -5.4 & 0.06 & 0.02\\
$m_2$ & 0.0 & 4.0 & -3.2 & -5.2 & 0.03 & 0.02
\end{tabular}
\caption{Expected log Bayes factors for models with different infection mechanisms, assuming diffuse prior distributions. Model $m_1$ is a standard SIR
model while $m_2$ has a modified infection mechanism of the form $\beta n^{-1} X(t) Y^p(t)$. Both models have $T_I \sim \mbox{Exp}(1)$. Each row gives parameter values
and results from 1000 simulated epidemics from the true model in which at least one new infection occurred.}
\end{table}

\section{Model selection given incomplete outbreak data}\label{Model selection given incomplete outbreak data}
We now consider the situation in which we observe removal times but not infection times, which in turn means that the
Bayes factors of interest are no longer analytically tractable. In this section we
describe how to apply the extended power posterior methods in section \ref{Power posteriors for models incorporating missing data} to
the model comparison scenarios in sections \ref{The SIR model with different infectious periods} and \ref{The SIR model with different infection mechanisms}.
In both cases, Algorithm \ref{algor1} requires an MCMC scheme that provides samples from the power posterior distribution for any given value of $t_j$. Since
the infection times are unobserved, these are included as additional components of the posterior distribution. Thus the required MCMC algorithm may be specified
by defining the updates for each of the model parameters, the details of which are given below.

We also briefly explore the performance of the power posterior methods and DIC$_6$, via simulation studies. It should be noted that such simulations are highly computationally expensive and time-consuming, since we require separate runs of an MCMC algorithm for every single $t_j$ value in the temperature ladder.
Throughout we set $c=5$ so that $t_j = (j/r)^5$. In all cases, the results are based on MCMC runs of 27,000 iterations of which the first 2,000 were discarded as burn-in, and then thinned by taking every 5th value. Convergence and mixing were assessed visually, and found to be satisfactory.

\subsection{The SIR model with different infectious periods} \label{Application of the updated power posterior and DIC$_6$ methods to the SIR model with different infectious periods}

Recall the models and notation from section \ref{The SIR model with different infectious periods}. The parameters $\beta$, $\gamma$ and $\delta$ are assigned independent exponential prior distributions $\mbox{Exp}(\lambda_\zeta)$, where $\zeta=\beta,\;\gamma,\;\delta$.
We assume {\em a priori} that the initial infection time satisfies $I_z = R_{min} - Y$, where $Y \sim \mbox{Exp}(\psi)$ and $R_{min} = \min \left\{ R_1, \ldots, R_{n_R} \right\}$, and that the initial infective $z$ is equally likely to be any of the $n_R$ infected individuals.

At temperature $t$, the full conditional power posterior distributions for $\beta, \gamma$ and $\delta$ are
\begin{equation*}
  \beta\vert t, \gamma, z, I_z, \bm{I},\bm{R}\sim \mbox{Gamma} \left(1 + t(n_R-1), \lambda_\beta + t n^{-1} A \right),
  \end{equation*}
\begin{equation*}
  \gamma\vert t, \beta, z, I_z, \bm{I},\bm{R}\sim \mbox{Gamma} \left(1 + n_R t, \lambda_\gamma + t \sum_{j=1}^{n_R}(R_j-I_j)\right),
  \end{equation*}
\begin{equation*}
  \delta\vert t, \beta, z, I_z, \bm{I},\bm{R}\sim \mbox{Gamma} \left(1 + t \alpha n_R, \lambda_\delta + t \sum_{j=1}^{n_R}(R_j-I_j)\right),
  \end{equation*}
and the full conditional power posterior density for $(\bm{I}, z, I_z)$ under model $m_2$  is given by
\begin{eqnarray*}
\pi(\bm{I}, z, I_z \vert t,\alpha, \beta, \delta, \bm{R})&\propto& \left\{\prod_{j=1,j\neq z}^{n_R} n^{-1}Y(I_j-) \times e^{ -\beta n^{-1} A} \right\}^t \\
&\times& \left\{\prod_{j=1}^{n_R}(R_j-I_j)^{\alpha-1}\times e^{-\delta \sum_{j=1}^{n_R}(R_j-I_j)}\right\}^t \times e^{\psi I_z},
\end{eqnarray*}
while the corresponding expression for model $m_1$ is obtained by setting $\alpha=1$ and $\delta = \gamma$. An MCMC algorithm to update the model parameters
and infection times then consists of (i) updating  $\beta, \gamma$ and $\delta$ according to their full conditional distributions, and (ii) updating infection
times using a suitable Metropolis-Hastings step as in \cite{oneill1999}; full details can be found in \cite{alharthi2016}.

The DIC$_6$ calculation involves two steps. An initial MCMC run provides point estimates of the model parameters, the identity of the initial infective and
the initial infection time. The model parameters and initial infective conditions are fixed for a second MCMC run in which the remaining infection times
are allowed to vary. From each run we also obtain an estimate of the posterior mean of the augmented log-likelihood, from which DIC$_6$ can be computed
according to (\ref{DIC_formula}).

\subsubsection{Simulation study}\label{Factors affecting the updated power posterior and DIC$_6$ performances}
\begin{figure} [h!t]
\centering
\begin{tabular}{c}
\includegraphics[scale=0.45,trim=0 20 0 20]{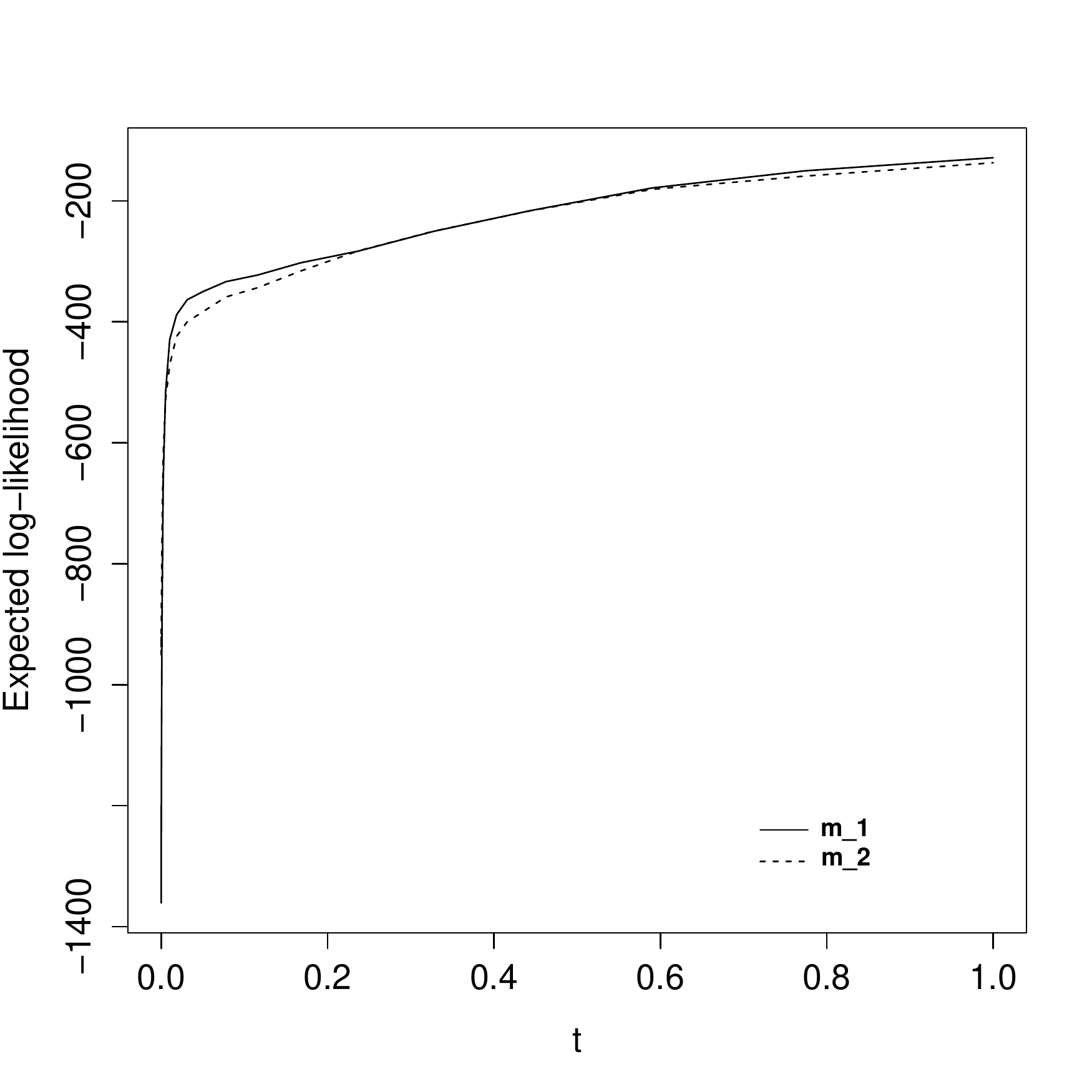}\\\\
\includegraphics[scale=0.45,trim=0 20 0 20]{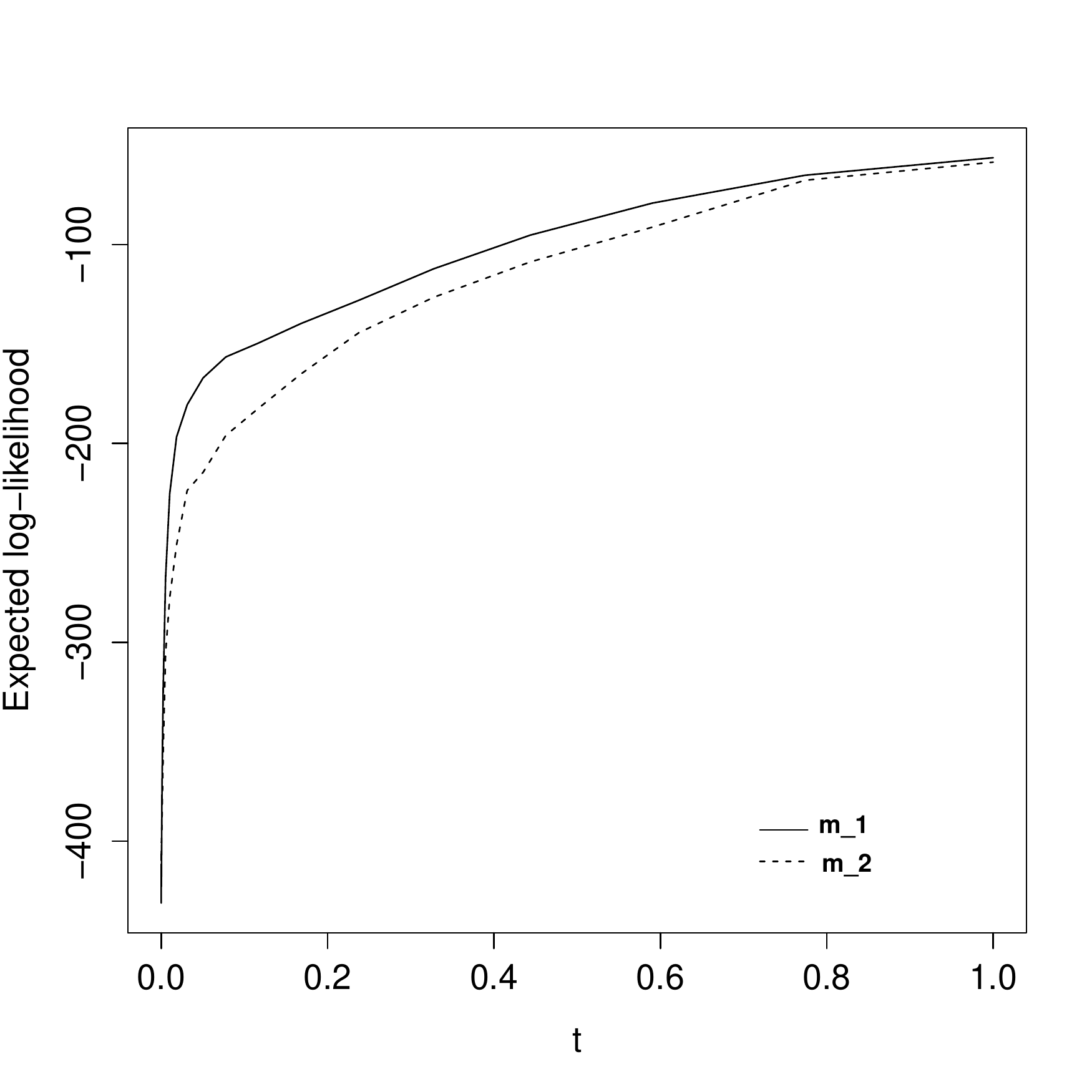}
\end{tabular}
\caption{Expected log-likelihood curves given data generated under the SIR model with $T_I\sim \mbox{Exp}(\gamma)$ ($m_1$, top plot) and under the SIR model with
$T_I\sim \mbox{Gamma}(\alpha=10, \delta)$ ($m_2$, bottom plot). In both cases $\beta$, $\gamma$ and $\delta$ were assigned independent $\mbox{Exp}(1)$ prior distributions.} \label{expectedL2}
\end{figure}

We now briefly assess the impact of the temperature ladder (i.e. the set of $t_j$ values in Algorithm \ref{algor1}), the within-model prior distribution, and
the size of the observed epidemic on the calculation of Bayes factors. We set $\psi=1$ in the prior distribution of $I_z$.
\begin{itemize}
\item \textbf{Length of temperature ladder}
\end{itemize}
Table \ref{table4.BF.DIC} shows the impact of the number of $t_j$ values, $r$, on the marginal likelihoods and corresponding Bayes factor $BF_{12}$. The results
are based on a single, but fairly typical, data set simulated from model $m_1$ with $N=30$, $\beta = 1$, $\gamma = 0.5$ and in which $n=22$ individuals were
infected in total. In model $m_2$, $\alpha = 10$. Two choices of prior distribution are illustrated. The results show that the estimates are
relatively insensitive to the value of $r$, although as expected more $t_j$ values are required as the prior distribution becomes more diffuse. Figure \ref{expectedL2}
shows typical expected log-likelihood curves, and illustrates the sharp change near $t=0$ which motivates the choice of temperature ladder $t_j = (j/r)^c$.
\begin{table} [h!t]
  \centering
\begin{tabular}{cccc}\hline
$r$ &$\log(\pi(\bm{R}\vert m_1))$  & $\log(\pi(\bm{R}\vert m_2))$  & $\log(BF_{12})$\\\hline
\multicolumn{4}{c}{$\beta, \gamma, \delta \sim \mbox{Exp}(1)$} \\\hline
$10$ & $-129.86  $ & $ -152.25 $ & $ 22.39$\\
$20$ & $ -130.01 $ & $ -150.87  $ & $20.86$\\
$40$ & $ -130.05$ & $ -151.15 $ & $ 21.11$\\
$100$ & $ -130.04  $ & $ -150.24 $ & $ 20.19 $\\\hline
\multicolumn{4}{c}{$\beta, \gamma, \delta \sim \mbox{Exp}(0.01)$ } \\\hline
$10$ & $ -131.97  $ & $ -150.71 $ & $ 18.74$\\
$20$ & $ -137.07  $ & $ -151.98 $ & $ 14.91$\\
$40$ & $ -137.49  $ & $ -153.04  $ & $ 15.56$\\
$100$ & $ -137.39  $ & $ -152.36 $ & $ 14.98$\\\hline
\end{tabular}
\caption{Estimates of $\log(\pi(\bm{R}\vert m_1))$, $\log(\pi(\bm{R}\vert m_2))$ and $\log(BF_{12})$ using data simulated from the standard SIR model with $T_I\sim \mbox{Exp}(\gamma)$ ($m_1$), while model $m_2$ has $T_I\sim \mbox{Gamma}(\alpha, \delta)$. Parameter values were $N=30$, $\beta = 1$ and $\gamma=0.5$, and $n_R=22$
individuals were infected. The parameters $\beta, \gamma$ and $\delta$ were assigned $\mbox{Exp}(1)$ (top table) and $\mbox{Exp}(0.01)$ (bottom table) prior distributions.} \label{table4.BF.DIC}
\end{table}

\begin{itemize}
\item \textbf{Choice of prior distribution}
\end{itemize}
It is well known that Bayes factors can exhibit strong dependence on the model parameter prior distributions. Here, we explore this issue via two simulated data sets
from the two models under consideration. In both cases the data set itself was fairly typical of epidemics that did not die out quickly. For model $m_1$ we set $\beta = 2$, $\gamma = 1$ and $N=50$ and obtained $n_R=41$ infected individuals. For model $m_2$ we set $\beta=2$, $\alpha = 10$, $\delta = 10$ and $N=30$, and obtained $n_R=22$.
Prior distributions for $\beta$, $\gamma$ and $\delta$ were set to be $\mbox{Exp}(1), \mbox{Exp}(0.1)$ and $\mbox{Exp}(0.01)$, for which we used $r$ values of 20, 20
and 40 respectively, inspired by our previous findings regarding the length of temperature ladder.

Results of the simulation study are summarised in Table \ref{table1.BF.DIC}. The expected log-likelihood curves for both SIR models are shown
in Figure \ref{expectedL2} in the case when the prior distributions are $\mbox{Exp}(1)$.
\begin{table} [h!t]
\centering
\begin{tabular}{ccccc}\hline
\multirow{2}{*}{Model} & \multirow{2}{*}{Prior} & \multirow{2}{*}{ $\log(BF_{12})$} & \multicolumn{2}{c}{ DIC$_6$ } \\\cline{4-5}
       &   &   & $m_1$& $m_2$\\\hline
$m_1$ & $\mbox{Exp}(1)$ & $22.48$ & $\bm{276.21}$ & $291.01$ \\
$m_1$ & $\mbox{Exp}(0.1)$ & $10.00$ & $261.25$ & $\bm{248.04}$ \\
$m_1$ & $\mbox{Exp}(0.01)$ & $9.27$ & $261.30$ & $\bm{247.73}$ \\\hline
$m_2$ & $\mbox{Exp}(1)$ & $14.53$ & $\bm{123.60}$ & $127.76$ \\
$m_2$ & $\mbox{Exp}(0.1)$ & $-0.03$ & $115.02$ & $\bm{105.07}$ \\
$m_2$ & $\mbox{Exp}(0.01)$ & $-0.89$ & $114.63$ & $\bm{102.71}$ \\\hline
\end{tabular}
\caption{Estimates of $\log(BF_{12})$ and DIC$_6$ using data simulated from the SIR model with $T_I\sim \mbox{Exp}(\gamma)$ ($m_1$; $N=50$, $\beta = 2$, $\gamma=1$
and $n_R=41$ infections) and the SIR model with $T_I\sim \mbox{Gamma}(\alpha, \delta)$ ($m_2$; $N=30$, $\beta=2$, $\alpha = \delta = 10$, and $n_R=22$). Bold values for DIC$_6$ indicate the preferred model. In all cases $\beta$, $\gamma$ and $\delta$ were assigned identical independent prior distributions as indicated.} \label{table1.BF.DIC}
\end{table}
When model $m_1$ is the true model, the results in Table \ref{table1.BF.DIC} indicate, in general, that $\log(BF_{12})$ values support the true model for all prior distributions with a noticeable decrease as the prior distribution becomes more diffuse. These findings are in harmony with the behaviour of the Bayes factor for complete data case described in section \ref{The SIR model with different infectious periods}. When model $m_2$ is the true model, the value of $\log(BF_{12})$ varies quite dramatically with different prior distributions. As the latter become more diffuse then $m_2$ is identified correctly, whereas using an $\mbox{Exp}(1)$ prior gives the opposite conclusion. There is some intuition to explain this conflict, as follows. The true value of $\delta$ parameter used in the simulation is $10$, so an $\mbox{Exp}(1)$ prior is a strong prior distribution which is in conflict with the data and in turn results in poor posterior mean estimates for both $\beta$ and $\delta$, namely $\hat{\beta}=1.286$ and $\hat{\delta}=6.629$. This in turn yields lower values of $\log(\pi(\bm{R}\vert m_2))$ and thus $m_2$ is not identified correctly. However, with $\mbox{Exp}(0.1)$ and $\mbox{Exp}(0.01)$ priors, the estimation was improved giving $\hat{\beta}=2.053, \hat{\delta}=11.490$ and $\hat{\beta}=2.184, \hat{\delta}=12.133$, respectively, and consequently the correct identification of model $m_2$ was obtained. These results highlight the
sensitivity of Bayes factors to prior distributions, but also suggest that in this situation diffuse priors are likely to be more appropriate.

The values of DIC$_6$ are also prior-dependent. More seriously, model $m_2$ is preferred as the prior distributions become more diffuse, which suggests that
DIC$_6$ is not a suitable tool for model discrimination in this setting.

\begin{itemize}
\item \textbf{Size of outbreak}
\end{itemize}
It is natural to suppose that, given a small epidemic outbreak, it is hard to effectively distinguish between competing models. Here we show that even small
outbreaks can be informative in the setting where we compare infectious period distributions.

\begin{figure} [h!t]
\centering
\begin{tabular}{c}
\includegraphics[scale=0.5,trim=0 20 0 20]{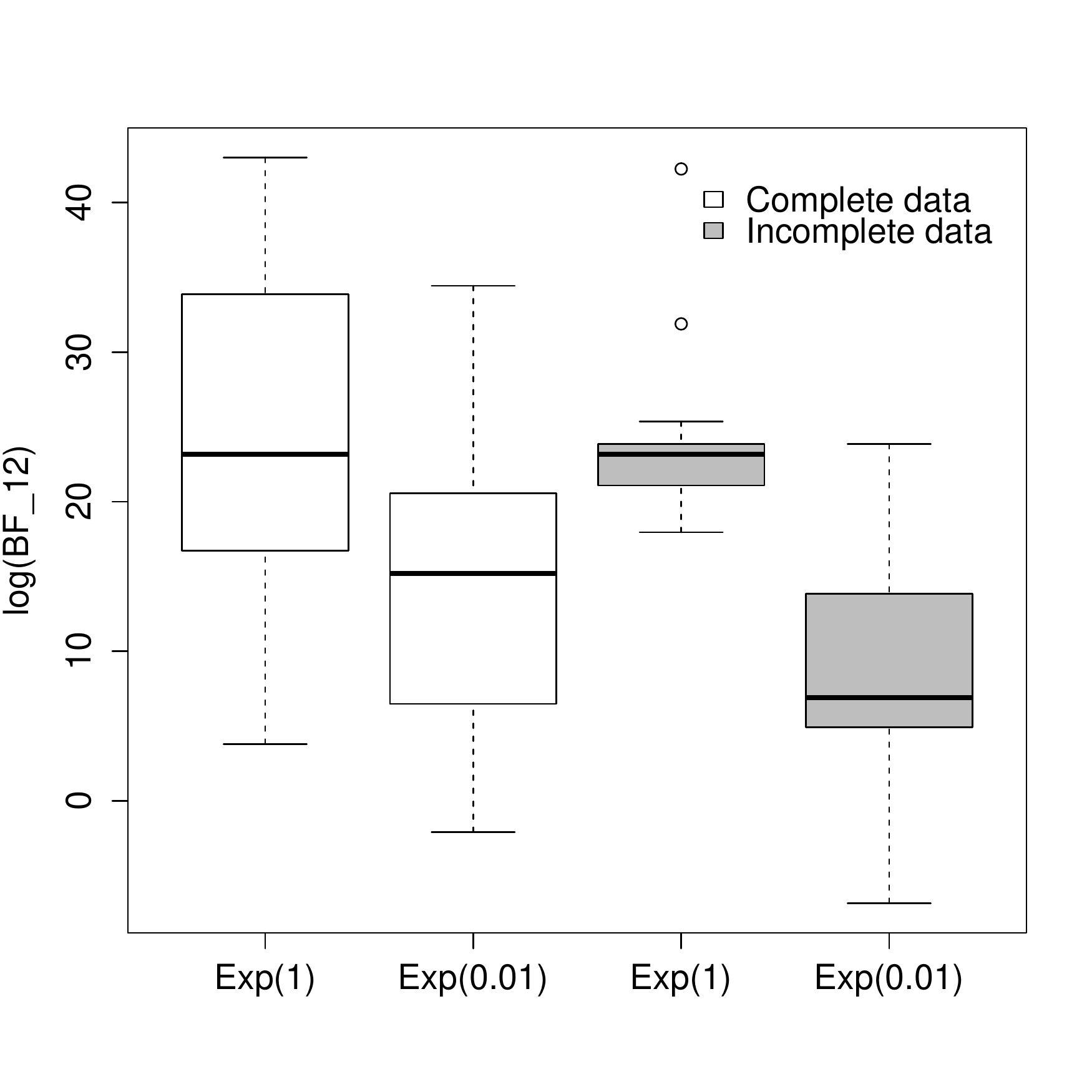}
\end{tabular}
\caption{Boxplots of $\log(BF_{12})$ values calculated using $20$ simulated data sets of $n_R=5$ removals each simulated under the SIR model with $T_I\sim \mbox{Exp}(\gamma)$ ($m_1$), while $m_2$ has $T_I\sim \mbox{Gamma}(\alpha=10, \delta)$. The $\log(BF_{12})$ values were computed using (\ref{BFlimitexpgam1}) and the missing-data power posterior method for complete and incomplete data respectively. The parameters $\beta, \gamma$ and $\delta$ were assigned two choices of prior distribution, namely $\mbox{Exp}(1)$ and $\mbox{Exp}(0.01)$.} \label{boxplot1}
\end{figure}

We simulated, under model $m_1$ with $N=50$, $\beta=1.15$ and $\gamma=1$, 20 datasets of size $n_R=5$ removals each.
For each data set we calculated $\log(BF_{12})$ for the complete data (infection and removal times), and for incomplete data (removal times), the latter using $r=20$, assuming that $\alpha=10$ in model $m_2$. Two different prior distributions were used for $\beta$, $\gamma$ and $\delta$, namely $\mbox{Exp}(1)$ and $\mbox{Exp}(0.01)$. The results are summarised Figure \ref{boxplot1}.

Interestingly, the results are not as one might expect with a small outbreak data set, giving decisive support to the true model $m_1$ under both complete and incomplete data. These findings suggest that, in this particular epidemic setting, a few infectious periods of infected individuals might be enough for the Bayes factor criterion to favour the SIR model with exponential infectious period over the SIR model with gamma infectious period. The impact of the parameter prior distributions is also evident which is in agreement with our previous findings. Note also that there is more variance in the 20 Bayes factors for complete than incomplete data. This is
essentially a consequence of the fact that calculating $BF_{12}$ for incomplete data involves averaging over the unobserved infection times, so the variability we
obtain from the original simulations is reduced.

\subsection{The SIR model with different infection mechanisms} \label{Application of the updated power posterior and DIC$_6$ methods to the SIR model with different infection mechanisms}
We now consider the models in section \ref{The SIR model with different infection mechanisms}, so that $m_1$ is the standard SIR model and $m_2$ has infection
rate $\beta n^{-1} X(t) Y^p(t)$. Prior distributions are assigned as in section \ref{Application of the updated power posterior and DIC$_6$ methods to the SIR model with different infectious periods}, and we also set $p\sim U(0, 0.5)$ {\em a priori}.

For model $m_2$ at temperature $t$, we have
\begin{equation*}
  \beta\vert t, \gamma, p, I_z, z, \bm{I}, \bm{R} \sim \mbox{Gamma} \left(1 + t(n_R -1),\lambda_\beta + t n^{-1} \int_{I_z}^{R_{n_R}} X(t)Y^p(t) dt \right),
  \end{equation*}
\begin{equation*}
  \gamma\vert t, \beta, p, I_z, z, \bm{I}, \bm{R} \sim \mbox{Gamma}\left(1 + n_R t, \lambda_\gamma + t \int_{I_z}^{R_{n_R}} Y(t)  dt \right).
  \end{equation*}
Conditional densities for $p$ and $\mathbf{I}$ are given by
\begin{equation*}
\pi(p\vert t, \beta, \gamma, I_z, z, \bm{I}, \bm{R})\propto \left \{ \left(\prod_{j=1, j \neq z}^{n_R} Y^{p}(I_j-)\right)\times \exp\left( -\beta n^{-1} \int_{I_z}^{R_{n_R}} X(t)Y^{p}(t) dt \right)\right\}^t,
\end{equation*}
\begin{eqnarray*}
\pi(\bm{I}\vert t, \beta, \gamma, p, I_z, z, \bm{R})&\propto& \left \{ \prod_{j=1, j \neq z}^{n_R} X(I_j-)Y^{p}(I_j-)\times \prod_{j=1}^{n_R} Y(R_j-)\right\}^t\\
&\times& \left \{\exp\left(-\int_{I_z}^{R_{n_R}}\left(\beta n^{-1}X(t)Y^{p}(t)+\gamma Y(t)\right)dt\right)\right\}^t.
 \end{eqnarray*}
The corresponding expressions for model $m_1$ can be obtained by setting $p=1$.

\subsubsection{Simulation study}\label{Factors influencing the updated power posterior and DIC$_6$ performances}
We consider the factors described in section \ref{Factors affecting the updated power posterior and DIC$_6$ performances}, and again set $\psi = 1$.

\begin{itemize}
\item \textbf{Length of the temperature ladder}
\end{itemize}
Table \ref{table6.BF.DIC} shows how estimates of marginal likelihoods and Bayes factors vary with $r$. The results are based on a simulation from model $m_2$ in
which $N=100$, $\beta=2$, $\gamma=0.2$ and $p=0.3$ which resulted in $n_R=87$ infections in total. Our findings are the same as those described in section
\ref{Factors affecting the updated power posterior and DIC$_6$ performances}, namely that as the prior distributions become more diffuse, more temperatures
are required to provide an accurate estimates.

\begin{table} [h!t]
  \centering
\begin{tabular}{cccc}\hline
$r$ &$\log(\pi(\bm{R}\vert m_1))$  & $\log(\pi(\bm{R}\vert m_2))$ & $\log(BF_{12})$\\\hline
\multicolumn{4}{c}{$\beta, \gamma \sim \mbox{Exp}(1)$ } \\\hline
$10$ & $ -113.35$ & $ -104.29$ & $ -9.06$\\
$20$ & $ -114.05$ & $ -106.07$ & $-7.99$\\
$40$ & $ -114.09$ & $ -105.76$ & $ -8.33$\\
$100$ & $ -113.80$ & $ -105.87$ & $ -7.93$\\\hline
\multicolumn{4}{c}{$\beta, \gamma \sim \mbox{Exp}(0.01)$ } \\\hline
$10$ & $ -81.087$ & $ -85.72$ & $ 4.63$\\
$20$ & $ -120.70$ & $ -111.03$ & $-9.67$\\
$40$ & $ -122.22$ & $ -111.59$ & $ -10.63$\\
$100$ & $ -122.28$ & $ -112.11$ & $ -10.17$\\\hline
\end{tabular}
\caption{Estimates of $\log(\pi(\bm{R}\vert m_1))$, $\log(\pi(\bm{R}\vert m_2))$ and $\log(BF_{12})$ using data simulated from the SIR model with modified infection
rate ($m_2$; $N=100$, $\beta=2$, $\gamma=0.2$, $p=0.3$ and $n_R=87$ infections), while $m_1$ is the standard SIR model. The parameters $\beta$ and $\gamma$ were assigned $\mbox{Exp}(1)$ (top table) and $\mbox{Exp}(0.01)$ (bottom table) prior distributions.} \label{table6.BF.DIC}
\end{table}
\begin{itemize}
\item \textbf{Choice of prior distribution}
\end{itemize}
We simulated one data set from each model, in both cases fairly typical. For model $m_1$ we set $N=100$, $\beta=0.5$, $\gamma=0.2$ and obtained $n_R=83$ infections.
For model $m_2$ we set $N=100$, $\beta=2.5$, $\gamma=0.2$ and $p=0.3$ and obtained $n_R=88$. Table \ref{table5.BF.DIC} shows estimates of $\log(BF_{12})$ and
DIC$_6$ under three choices of prior distribution, where we used $r$ values of 20, 40 and 40 for $\mbox{Exp}(1), \mbox{Exp}(0.1)$ and $\mbox{Exp}(0.01)$ prior
distribution, respectively.

Results of simulations are displayed in Table \ref{table5.BF.DIC}. Again there is evidence of sensitivity of $BF_{12}$ to the choice of prior, although the results
themselves show that the correct model is identified in all cases. Furthermore DIC$_6$ also performs well in this setting.

\begin{figure} [h!t]
\centering
\begin{tabular}{c}
\includegraphics[scale=0.45,trim=0 20 0 20]{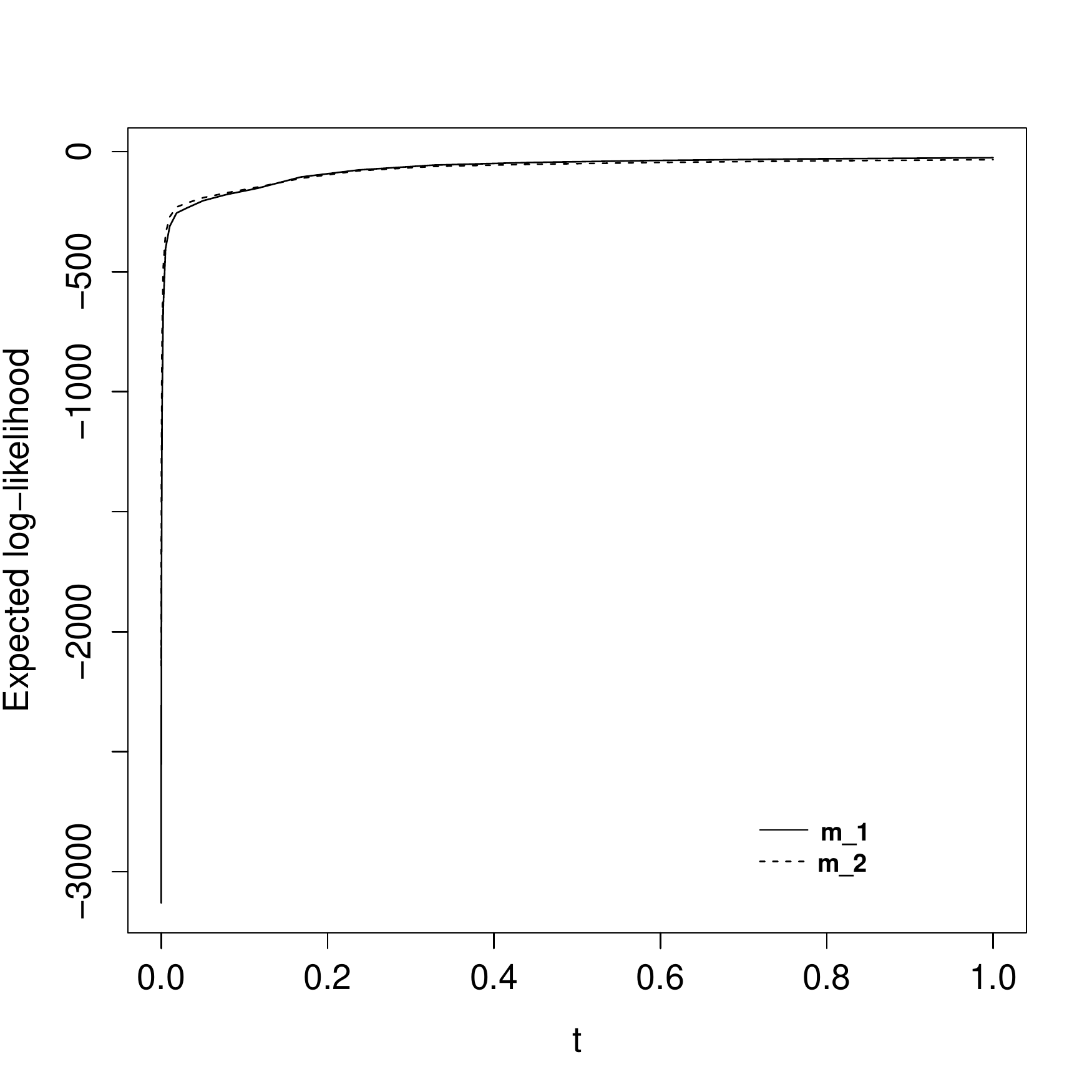}\\\\
\includegraphics[scale=0.45,trim=0 20 0 20]{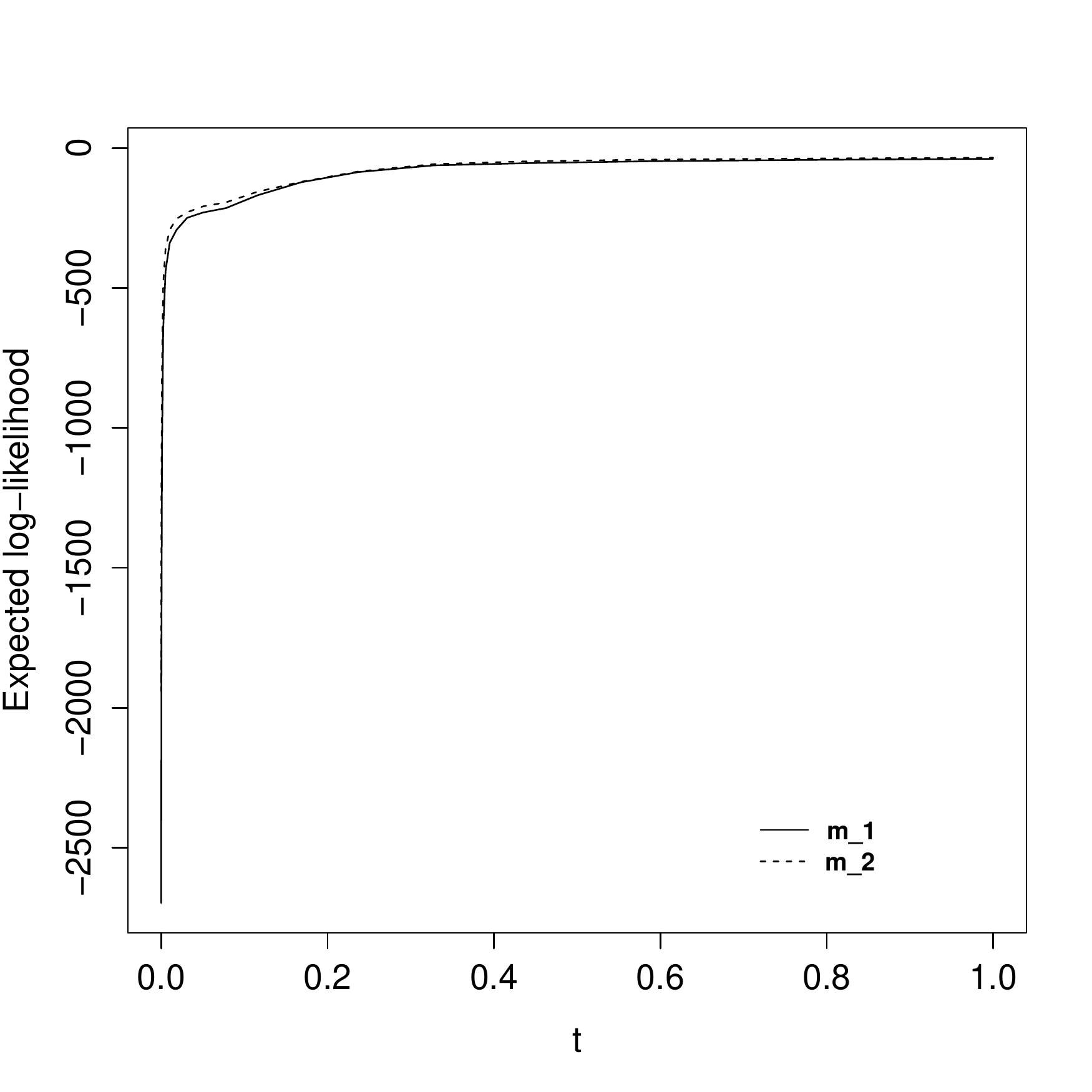}
\end{tabular}
\caption{Expected log-likelihood curves given data generated under the standard SIR model ($m_1$, top plot) and under the SIR model with modified infection rate
($m_2$, bottom plot). In both cases $\beta$ and $\gamma$ are assigned independent $\mbox{Exp}(1)$ prior distributions.} \label{expectedL1}
\end{figure}

Figure \ref{expectedL1} shows plots of the expected log-likelihoods against the temperature $t$ for the two simulated data sets. In contrast to Figure \ref{expectedL2},
here the curves for $m_1$ and $m_2$ are much closer together, indicating that it is harder to effectively discriminate between models with different infection
mechanisms than with different infectious period, at least for the settings we have considered.
\begin{table} [h!t]
  \centering
\begin{tabular}{ccccc}\hline
\multirow{2}{*}{Model} & \multirow{2}{*}{Prior} & \multirow{2}{*}{ $\log(BF_{12})$} &\multicolumn{2}{c}{ DIC$_6$ } \\\cline{4-5}
       &   &   & $m_1$& $m_2$\\\hline
$m_1$ & $\mbox{Exp}(1)$ & $3.49$ & $\bm{51.49}$ & $68.89$ \\
$m_1$ & $\mbox{Exp}(0.1)$ & $1.94$ & $\bm{52.08}$ & $68.09$ \\
$m_1$ & $\mbox{Exp}(0.01)$ & $2.10$ & $\bm{51.74}$ & $68.66$ \\\hline
$m_2$ & $\mbox{Exp}(1)$ & $-8.03$ & $76.57$ & $\bm{68.57}$ \\
$m_2$ & $\mbox{Exp}(0.1)$ & $-9.79$ & $76.98$ & $\bm{69.94}$ \\
$m_2$ & $\mbox{Exp}(0.01)$ & $-10.70$ & $77.37$ & $\bm{69.60}$ \\\hline
\end{tabular}
\caption{Estimates of $\log(BF_{12})$ and DIC$_6$ using data simulated from the standard SIR model ($m_1$;$N=100$, $\beta=0.5$, $\gamma=0.2$ and $n_R=83$) and the SIR model with modified infection rate ($m_2$; $N=100$, $\beta=2.5$, $\gamma=0.2$, $p=0.3$ and $n_R=88$).
Bold values for DIC$_6$ indicate the preferred model. In all cases $\beta$, $\gamma$ and $\delta$ were assigned identical independent prior distributions as indicated.} \label{table5.BF.DIC}
\end{table}

\begin{itemize}
\item \textbf{Size of outbreak}
\end{itemize}
We consider two scenarios for model $m_1$, corresponding to small and large epidemics. We simulated 20 epidemics for each, with the same number of removals $n_R$. In the first scenario $N=50$, $\beta=1.15$, $\gamma=1$ and $n_R=7$. In the second $N=50$, $\beta=2$, $\gamma=1$ and $n_R=42$. We set $p=0.3$ in model $m_2$. Bayes factor
calculations were carried out using $r=20$, and under two different prior distribution assumptions for $\beta$ and $\gamma$.

\begin{figure} [h!t]
\centering
\begin{tabular}{c}
\includegraphics[scale=0.5,trim=0 20 0 20]{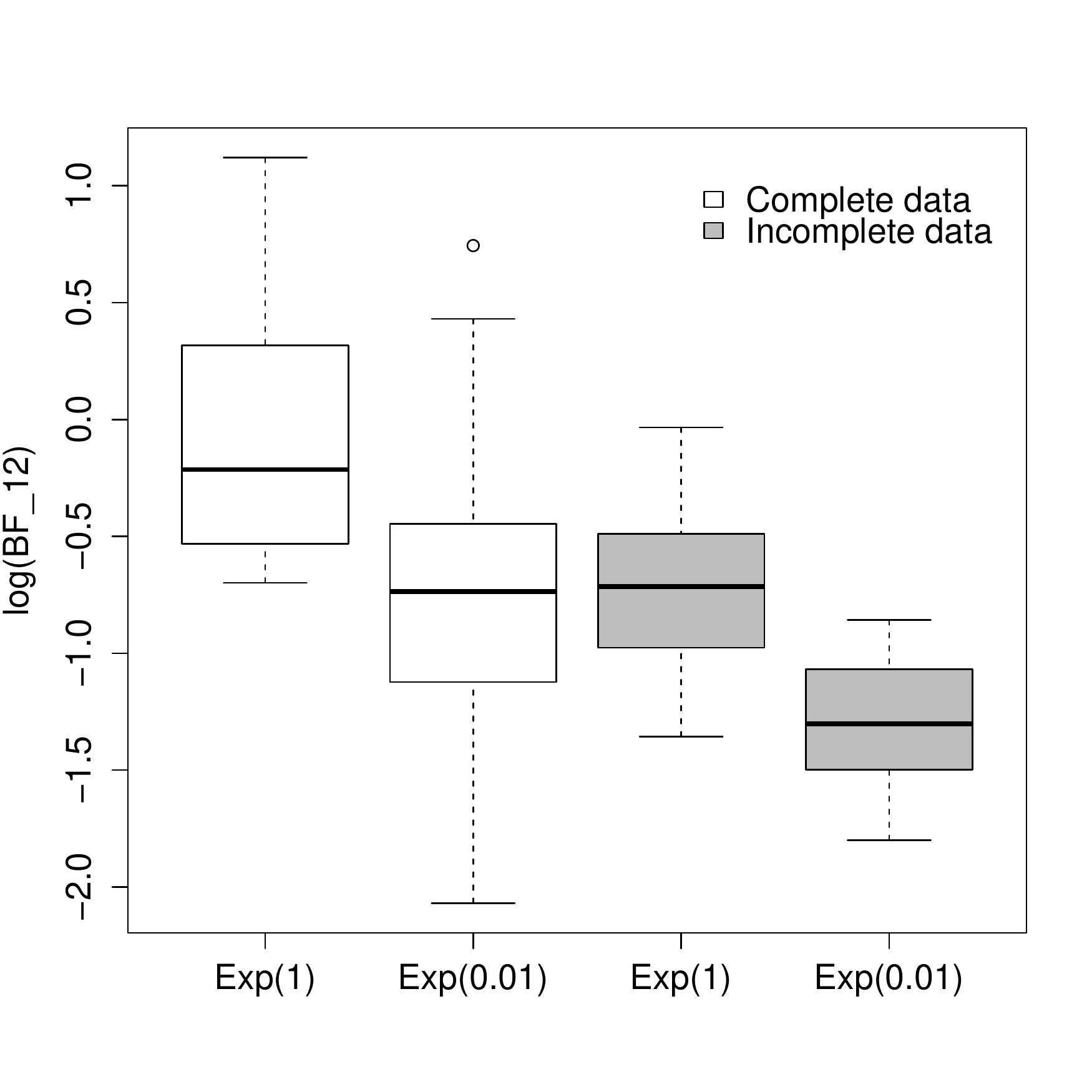}
\end{tabular}
\caption{Boxplots of $\log(BF_{12})$ values calculated using $20$ simulated data sets of $n_R = 7$ removals each simulated under the standard SIR model ($m_1$),
while $m_2$ has modified infection rate $\beta n^{-1} X(t) Y^p(t)$, $p=0.3$.
The $\log(BF_{12})$ values were computed using (\ref{true.BFstandvsmod}) and the missing-data power posterior method for complete and incomplete data respectively.
The parameters $\beta, \gamma$ and $\delta$ were assigned two choices of prior distribution, namely $\mbox{Exp}(1)$ and $\mbox{Exp}(0.01)$.} \label{boxplot2}
\end{figure}

\begin{figure} [h!t]
\centering
\begin{tabular}{c}
\includegraphics[scale=0.5,trim=0 20 0 20]{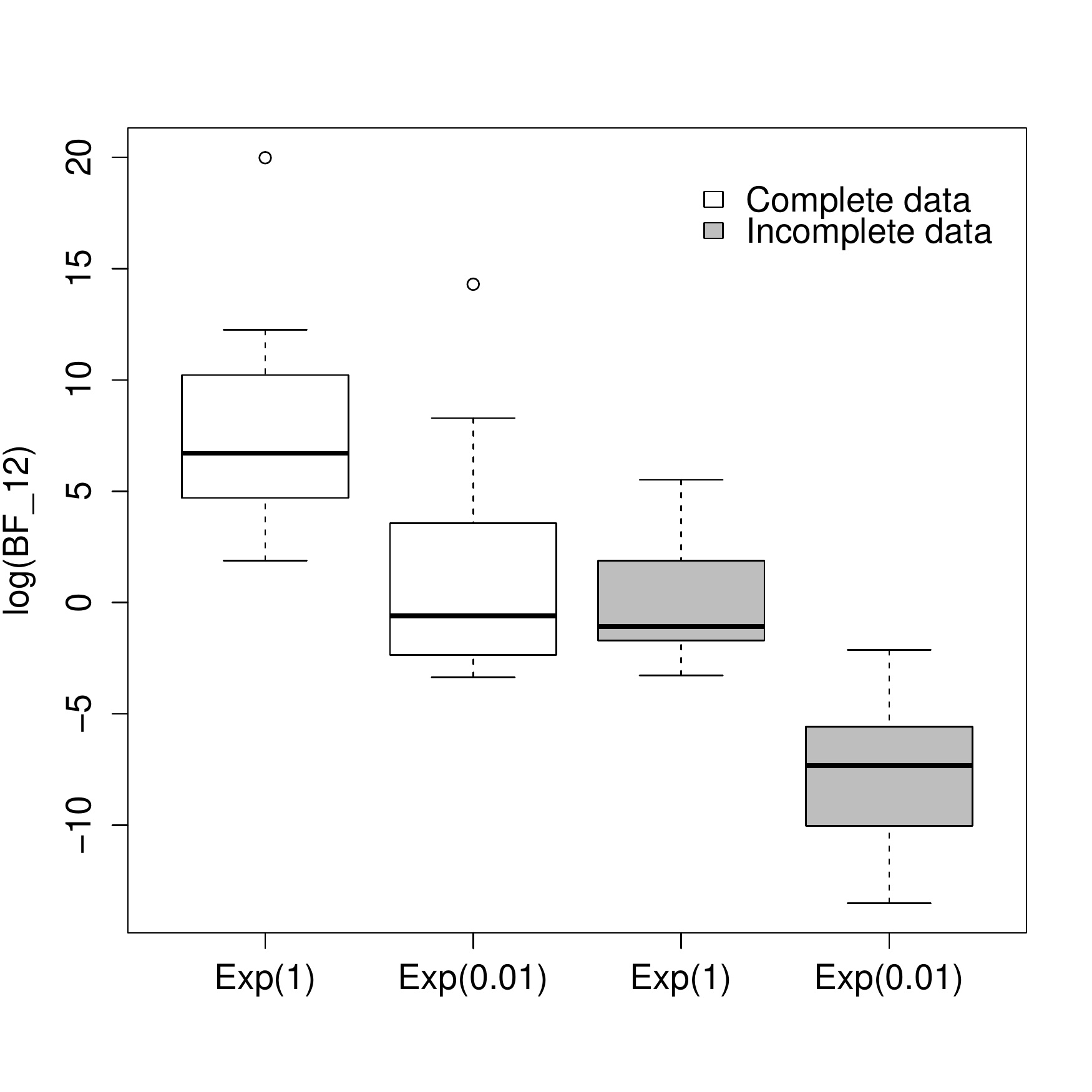}
\end{tabular}
\caption{Boxplots of $\log(BF_{12})$ values calculated using $20$ simulated data sets of $n_R=42$ removals each simulated under the standard SIR model ($m_1$),
while $m_2$ has modified infection rate $\beta n^{-1} X(t) Y^p(t)$, $p=0.3$.
The $\log(BF_{12})$ values were computed using (\ref{true.BFstandvsmod}) and the missing-data power posterior method for complete and incomplete data respectively.
The parameters $\beta, \gamma$ and $\delta$ were assigned two choices of prior distribution, namely $\mbox{Exp}(1)$ and $\mbox{Exp}(0.01)$.}
  \label{boxplot3}
\end{figure}

Figures \ref{boxplot2} and \ref{boxplot3} illustrate the results. In contrast to the comparison of infectious period distributions, here we see that small outbreaks
may not be sufficient to differentiate between models, which again agrees with our earlier findings that this is an inherently harder problem requiring more data.
However, even larger outbreaks appear problematic. The most likely reason for this is that the key to differentiating between $m_1$ and $m_2$ is the number of
infected individuals present in the population at the time of each infection, as discussed in section \ref{The SIR model with different infection mechanisms},
as opposed to the outbreak size itself. This suggests that data from epidemics that grow quickly would be required to distinguish the competing models, at least
for moderate population sizes.

\begin{figure} [h!t]
\centering
\begin{tabular}{c}
\includegraphics[scale=0.4,trim=0 20 0 20]{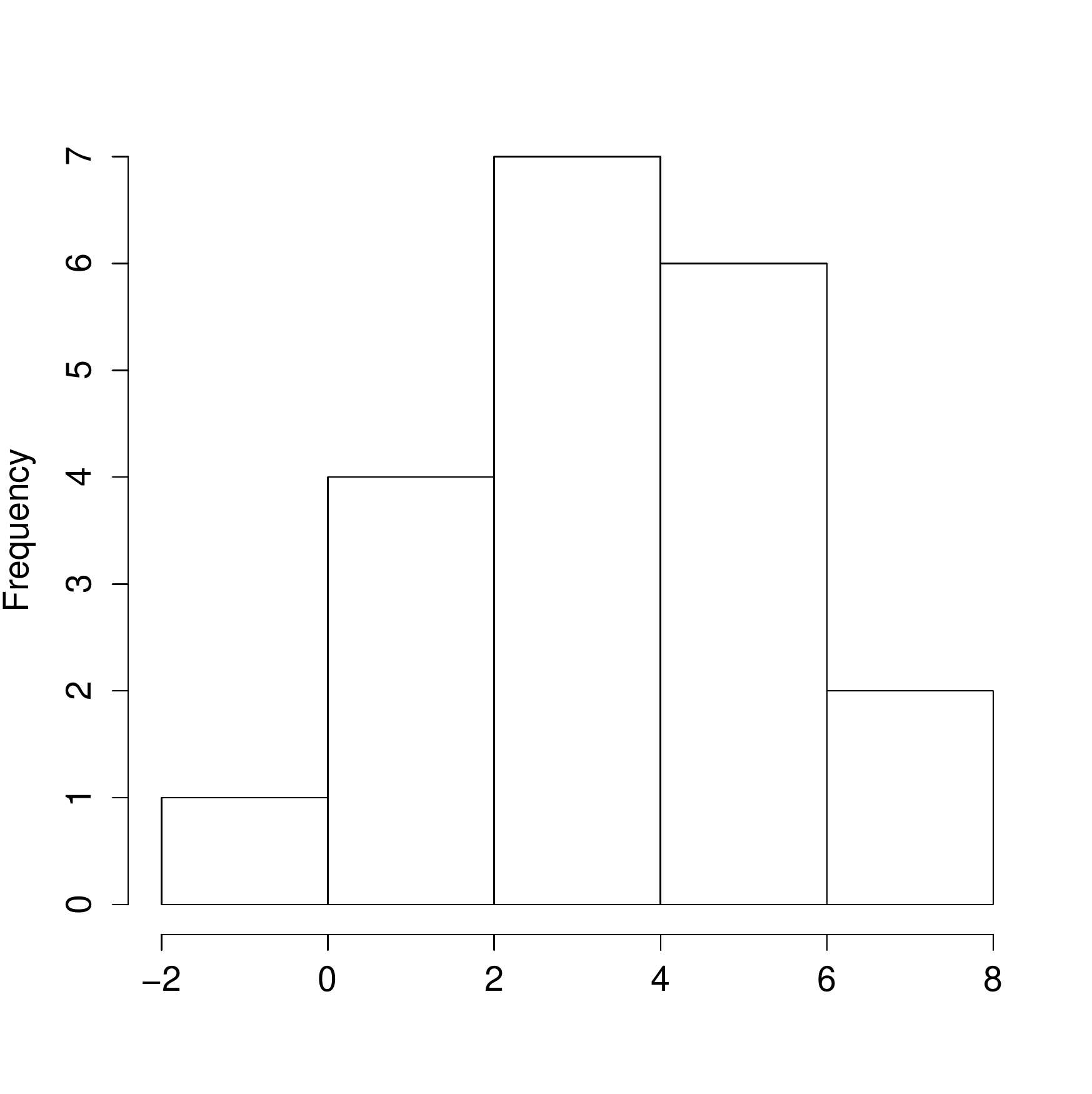}\\
$\log(BF_{12})$
\end{tabular}
\caption{Histogram of the estimated $\log(BF_{12})$ values obtained via the missing-data power posterior approach using $20$ simulated data sets of $n_R=47$ removal times each simulated under the standard SIR model ($m_1$), while $m_2$ has modified infection rate $\beta n^{-1} X(t) Y^p(t)$, $p=0.3$.
The parameters $\beta$ and $\gamma$ were assigned independent $\mbox{Exp}(1)$ prior distributions.} \label{histgram of log.BF12}
\end{figure}

To explore this further, we simulated 20 outbreaks of size $n_R=47$ from model $m_1$ with $N=50$, $\beta =5$ and $\gamma=1$. Thus $\beta/\gamma = 5$, in contrast
to the value of $2$ shown in Figure \ref{boxplot3}. Figure \ref{histgram of log.BF12} shows the resulting histogram of $\log(BF_{12})$ values from which we see that the evidence in favour of the true model $m_1$ is much clearer.

\subsection{Abakaliki smallpox data} \label{Application of the updated power posterior and DIC$_6$ methods to the Abakaliki smallpox data}
We now briefly consider a widely-studied temporal data set obtained from a smallpox outbreak that took place in the Nigerian town of Abakaliki in 1967.
The outbreak resulted in 32 cases, 30 of whom were members of a religious organisation whose 120 members refused vaccination. Numerous authors have considered
these data by focussing solely on the 30 cases among the population of 120, and the time series of symptom-appearance times which in our notation is given by
\begin{eqnarray*}
\bm{R}^{obs} &=& (0, 13, 20, 22, 25, 25, 25, 26, 30, 35, 38, 40, 40, 42, 42, 47, 50, 51, 55, 55, 56, \\
 & &57,58, 60, 60, 61, 66, 66, 71, 76),
\end{eqnarray*}
where to set a time scale we set $R_1 = 0$. In fact, the original data set includes far more information, particularly on the physical locations of the cases and the other members of their households. Analyses of this full data set can be found in \cite{eichner2003} and \cite{stockdale2017}.

Here our purpose is to illustrate our model comparison methods, and so we only consider the partial data set, specifically assuming that the 30
symptom-appearance times correspond to removals in an SIR model. Several authors have previously considered departures from the standard SIR model
for these data, and in particular both \cite{beckeryip1989} and \cite{Xia2015} considered models in which the infection rate was allowed to vary through
time. Here we address this issue by comparing the standard SIR model, $m_1$, with a model ($m_2$) in which the infection rate parameter $\beta$ is replaced by
$\beta(t)=\beta n^{-1} e^{-bt}$, with $b > 0$ a model parameter to be estimated from the data. It is straightforward to modify our methods to this situation.

\begin{figure} [h!t]
\centering
\includegraphics[scale=0.7,trim=0 20 0 20]{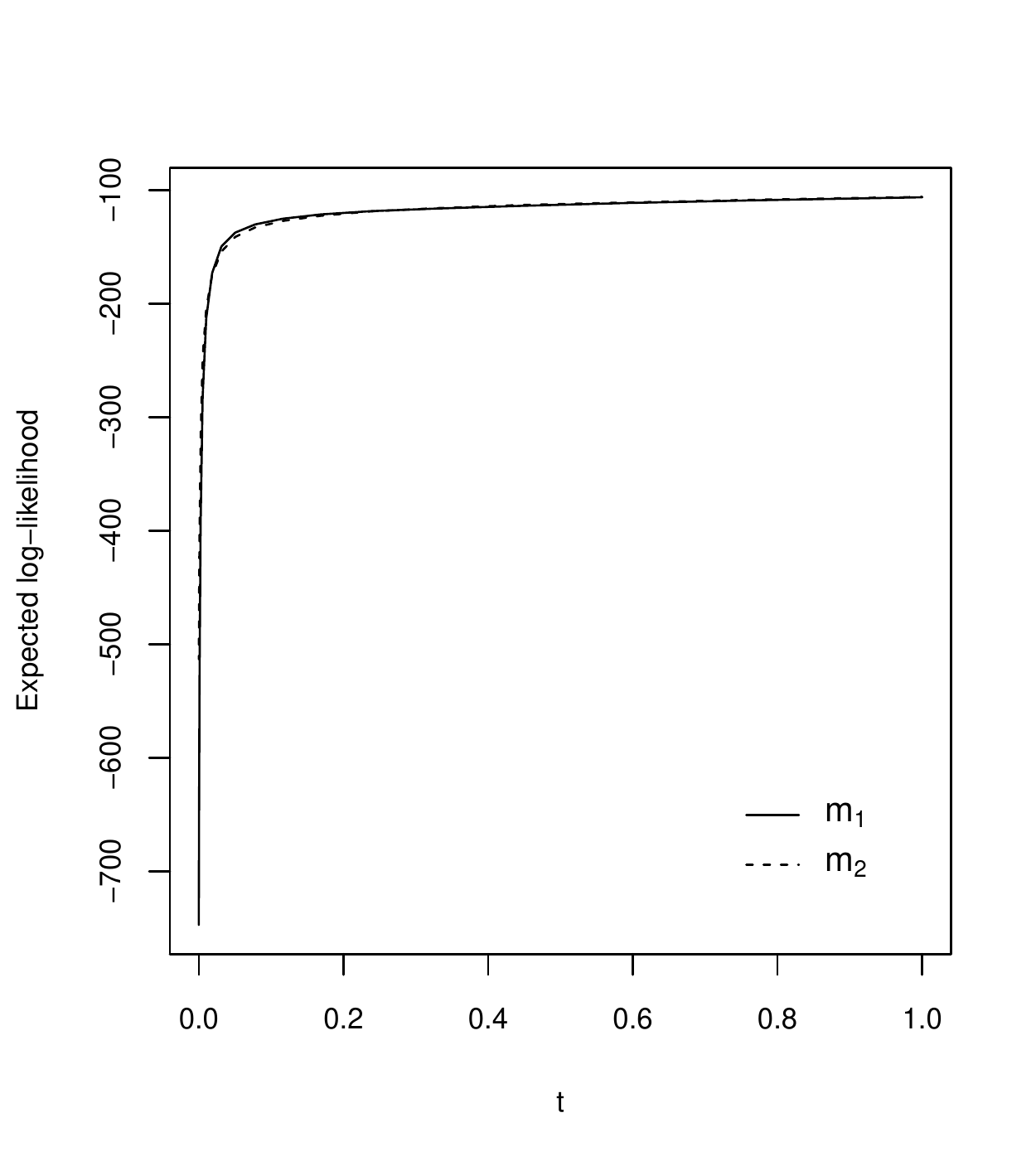}
  \caption{Smallpox data: expected log-likelihood curves calculated using the extended power posterior approach when $\beta, \gamma \sim \mbox{Exp}(1)$ {\em a priori.}} \label{smallpoxEloglike}
\end{figure}

Results are presented in Table \ref{table.smallpox.BF.DIC} for different prior distributions for $\beta$ and $\gamma$ for both SIR models, where $b$ and $R_1-I_1$ are assigned $\mbox{Exp(1)}$ prior distributions throughout. Figure \ref{smallpoxEloglike} shows plots of the expected log-likelihoods against the temperature $t$ when the prior distribution is $\mbox{Exp}(1)$, using a temperature ladder with $t_j = (j/r)^5$, $j=0, \ldots, 20$ and $r=20$.
\begin{table} [h!t]
  \centering
\begin{tabular}{cccc}\hline
 \multirow{2}{*}{Prior} & \multirow{2}{*}{ $\log(BF_{12})$} &\multicolumn{2}{c}{ DIC$_6$ } \\\cline{3-4}
   &   & $m_1$& $m_2$\\\hline
 $\mbox{Exp}(1)$ & $-0.51$ & $-105.8$ & $-105.9$ \\
 $\mbox{Exp}(0.01)$ & $-0.86$ & $-105.8$ & $-105.8$ \\\hline
\end{tabular}
\caption{Results of $\log(BF_{12})$ and DIC$_6$ for smallpox data based on the standard SIR model ($m_1$) and the modified SIR model ($m_2$). For each case, the parameters $\beta$ and $\gamma$ were assigned prior distributions as indicated.} \label{table.smallpox.BF.DIC}
\end{table}

\begin{figure} [h!t]
\centering
\includegraphics[scale=0.6,trim=0 20 0 20]{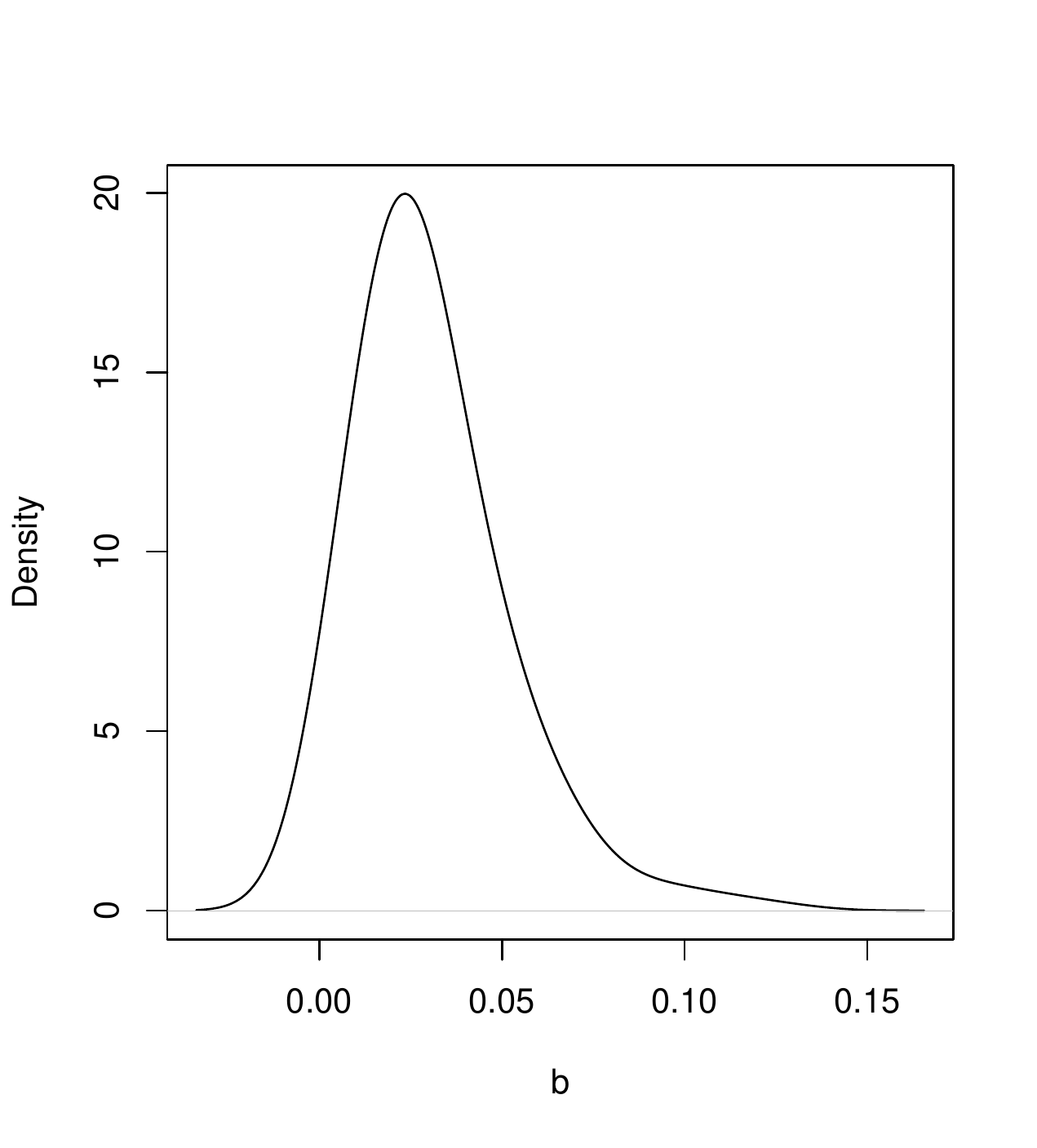}
  \caption{Smallpox data: posterior density of $b$, when $\beta, \gamma \sim \mbox{Exp}(1)$ {\em a priori.}} \label{post.b}
\end{figure}

From Table \ref{table.smallpox.BF.DIC}, there is clearly little to choose between the two models. This comparison is borne out from parameter estimation from
model $m_2$, specifically Figure \ref{post.b} which shows that $b$ is close to zero. These findings are in keeping with those in \cite{xu2016}, in which
a Bayesian nonparametric time-varying estimate of the infection rate parameter was found to be fairly close to constant over time.

\subsection{Hagelloch measles data} Our second data set describes an historical outbreak of measles in 1861 in the
German village of Hagelloch, as described in \cite{Pfeilsticker1861}. The outbreak was very severe, as every
one of 188 individuals deemed to be susceptible became infected, these individuals all being children.
These data have been considered by a number of authors \citep[see e.g][and references therein]{britton2011},
and were specifically analysed in a model choice context in \cite{neal2004}, where the authors used
reversible-jump MCMC methods to evaluate Bayes factors for a number of competing models. In view of our
earlier discussion on the possible sensitivity of Bayes factors to within-model prior distributions, it
is natural to explore this further for the Hagelloch data.

The data themselves are unusually detailed, consisting for each case of the name, age, sex, date of symptom
onset, date of rash onset, class of child in the village school, date of death if this occurred, location
of the child's home and several other covariates. We shall adopt the transmission models described in \cite{neal2004},
defined as follows.

Each individual belongs to a household and the community at large, and either attends school or is of pre-school age. If an individual becomes
infected, they first undergo a symptom-free infectious period $T_S$ which is assumed to be one of two models: either (i) a fixed length of one day, so $T_S = 1$,
or (ii) $T_S \sim Gamma(30, \delta)$. At the end of this period, the individual displays symptoms and subsequently develops a rash.
The dates of both symptom and rash appearance are given by the data and so we do not model the time between these events. Following the rash appearance,
the individual is assumed to be removed three days later, unless they die first (as indicated in the data). The infectious period is assumed to start upon initial
infection, and finish at either removal or death.

While infectious, individual $i$ has infectious contacts with susceptible individual $j$ at rate $\alpha_{ij}$ where $\alpha_{ij}$ depends on the relationship of $i$ and $j$. \cite{neal2004} first considered a model in which
$$\alpha_{ij} = \beta_{H}\, 1_{\{\rho(i,j)=0\}} + \beta^1_{C}\, 1_{\{L_i=L_j=1\}} + \beta^2_{C}\, 1_{\{L_i=L_j=2\}} + \beta_{G} \exp\{-\theta \rho(i,j) \}$$
where (i) $1_A$ denotes the indicator function of the event $A$, (ii) $\rho(i,j)$ denotes the Euclidean distance between the households of individuals $i$ and $j$, (iii) $L_i$ denotes the school classroom (either 1 or 2) which individual $i$ belongs to, and (iv) $L_i = 0$ if individual $i$ is of pre-school age. The non-negative parameters $\beta_H$, $\beta_C^1$ and $\beta_C^2$ denote the within-household, within-classroom 1 and within-classroom 2 infection rates respectively. Finally, $\beta_G$ denotes a global infection rate while $\theta$ governs the extent to which distance between individuals reduces this infection rate. \cite{neal2004} called the model described above the {\em full model}, denoted by $M$.

In order to investigate the relative importance of different transmission routes,
\cite{neal2004} considered four other models, each of which is a simplified version of the full model. Specficially, they set each of the parameters $\theta, \beta_H, \beta_C^1$ and $\beta_C^2$ to zero in turn in $M$, yielding models $M[\theta]$, $M[\beta_H]$, $M[\beta_C^1]$ and $M[\beta_C^2]$, say, respectively. Thus for example, $M[\beta_H]$ assumes that
$$\alpha_{ij} = \beta^1_{C}\, 1_{\{L_i=L_j=1\}} + \beta^2_{C}\, 1_{\{L_i=L_j=2\}} + \beta_{G} \exp\{-\theta \rho(i,j) \}.$$
Finally, for each parameter $\psi$ define $BF[\psi] = BF_{M,M[\psi]}$, so for example $BF[\theta] = BF_{M,M[\theta]}$ is the Bayes factor for the full model relative
to the model in which $\theta=0$.

We employed an MCMC algorithm which targets the joint power posterior distribution of the parameters $(\theta, \beta_H, \beta_C^1, \beta_C^2$ and  $\beta_G$ when
$T_S=1$, and additionally $\delta$ and the unknown infection times when $T_S \sim \mbox{Gamma}(30, \delta)$) given the observed data. None of the full conditional power posterior distributions for the model parameters are standard, and so parameters were updated using Metropolis-Hastings steps. The variances were tuned manually to achieve an acceptance rate approximately between 25\% and 40\%. To estimate the marginal likelihood via the power posterior method we used a temperature ladder such that $t_j = (j/r)^5$, where $j=0,\ldots, 100$ and $r=100$. Convergence and mixing was assessed visually, and found to be satisfactory.

Results are given in Table \ref{tab:logBF_Hagelloch} for different prior distributions for the model parameters. In \cite{neal2004}, an Exp(0.1) prior distribution
is assigned to the parameter $\theta$ and an Exp(10) to the parameters $\beta_H, \beta_C^1, \beta_C^2$ and $\beta_G$. We also consider two alternative sets of prior distributions, specifically assigning an Exp(1) or an Exp(0.1) distribution to {\em all} the parameters in each model.

The results illustrate a certain amount of sensitivity to both the $T_S$ model and the chosen prior distributions. Regarding the former, the posterior mean of $\delta$
was found to be around 4, so that when infection times are not assumed to be known the length of the symptom-free period was estimated to be around 7 days. This is in
stark contrast to the assumption that $T_S$ is one day, and goes some way to explaining the differences between the Bayes factors for the two model scenarios.
Note also that our estimate of $\delta$ is almost certainly connected to the fact that measles actually has a latent period of around 7-10 days, a feature missing from
the SIR model we consider; roughly speaking, increasing the infectious period length is one way to account for the time taken for the whole outbreak. We also briefly
investigated what happens if $T_S$ was set equal to integer values from 2 to 7 days, and found that the average posterior log-likelihood increased accordingly, which also supports the view that the $T_S=1$ model is less appropriate for these data than $T_S \sim \mbox{Gamma}(30, \delta)$.

Regarding the Bayes factors in Table \ref{tab:logBF_Hagelloch}, the only robust conclusion appears to be that there is strong evidence that $\beta_C^1 > 0$, itself a
key finding from \cite{neal2004}. For the $T_S \sim \mbox{Gamma}(30, \delta)$ model there is also strong support for the hypotheses that $\beta_C^2 > 0$ and that $\theta=0$. These findings make sense, since as illustrated in \cite{britton2011} it is clear from the raw data that the epidemic moves through the two school classes in turn, suggesting that classroom transmission is important, while there is no apparent evidence of purely spatial transmission.

\begin{table}[H]
\centering
\begin{tabular}{cccccc}
\hline
$T_S = 1$ & Priors      & $\log( BF[\theta])$ & $\log(BF[\beta_H])$ & $\log(BF[\beta_C^1])$ & $\log(BF[\beta_C^2])$ \\ \hline
&As Neal and Roberts     & -0.03                 & 5.77                & 33.46               & 6.72                \\
&Exp(1)               & 0.78                & 9.50                & 30.87               & 3.82                \\
&Exp(0.1)             & -0.64                 & 0.99                & 28.62               & 1.89 \\ \hline
\hline
$T_S \sim \mbox{Gamma}(30, \delta)$ & Priors      & $\log( BF[\theta])$ & $\log(BF[\beta_H])$ & $\log(BF[\beta_C^1])$ &$\log(BF[\beta_C^2])$ \\ \hline
&As Neal and Roberts     & -2.72                 & 1.73                & 70.97              & 12.71 \\
&Exp(1)               & -10.56                & 18.05               & 89.77              & 12.79 \\
&Exp(0.1)             & -28.34                & -8.23                 & 56.77           & 20.93 \\ \hline
\end{tabular}
\caption{Estimates of $\log$ Bayes Factors for the Hagelloch data for different models for the symptom-free period $T_S$ and different within-model prior assumptions.}
\label{tab:logBF_Hagelloch}
\end{table}

\section{Conclusions}
We have described a method for computing Bayes factors for epidemic models, specifically by utilising an extension of the power
posterior method to accommodate missing data of a certain kind. The methods appear to work reasonably well in practice. Although
we have focussed on single population SIR models, the methods can be applied to more complex epidemic models, as illustrated by
our examples featuring the Abakaliki smallpox data and the Hagelloch measles data.

In common with the original power posterior method itself, the main drawback with our methods is one of computational efficiency,
specifically that it can be quite time-consuming to perform the numerous MCMC runs required for the method. However, the
procedure is relatively easy to implement, and lends itself to parallel computation.

We have also demonstrated that Bayes factors can be obtained analytically for situations where infection times are known. This in
turn enables us to explore the impact of different prior assumptions analytically. Although the infection process is not usually
observed in real-life settings, if the infectious period does not exhibit that much variation then we can approximate it by a fixed
value. Under this assumption the infection times would be known, which could enable explicit evaluation of Bayes factors for comparing
different possible infection mechanisms or routes of infection, depending on the problem in question.

\bibliographystyle{ba}
\bibliography{References}

\end{document}